\documentclass[a4paper,11pt]{article}
\pdfoutput=1 

\usepackage{jcappub} 

\usepackage[T1]{fontenc} 
\usepackage{verbatim}

\def\setsymbol#1#2{\expandafter\def\csname #1\endcsname{#2}}
\def\getsymbol#1{\csname #1\endcsname}

\def\Planck{\textit{Planck}}

\newbox\tablebox    \newdimen\tablewidth
\def\leaderfil{\leaders\hbox to 5pt{\hss.\hss}\hfil}
\def\tablenote#1 #2\par{\begingroup \parindent=0.8em
    \abovedisplayshortskip=0pt\belowdisplayshortskip=0pt
    \noindent
    $$\hss\vbox{\hsize\tablewidth \hangindent=\parindent \hangafter=1 \noindent
    \hbox to \parindent{$^#1$\hss}\strut#2\strut\par}\hss$$
    \endgroup}

\def\L2{\ifmmode L_2\else $L_2$\fi}

\def\DeltaT{\ifmmode \Delta T\else $\Delta T$\fi}
\def\deltat{\ifmmode \Delta t\else $\Delta t$\fi}
\def\fknee{\ifmmode f_{\rm knee}\else $f_{\rm knee}$\fi}
\def\Fmax{\ifmmode F_{\rm max}\else $F_{\rm max}$\fi}
\def\solar{\ifmmode{\rm M}_{\mathord\odot}\else${\rm M}_{\mathord\odot}$\fi}
\def\Msolar{\ifmmode{\rm M}_{\mathord\odot}\else${\rm M}_{\mathord\odot}$\fi}
\def\Lsolar{\ifmmode{\rm L}_{\mathord\odot}\else${\rm L}_{\mathord\odot}$\fi}
\def\inv{\ifmmode^{-1}\else$^{-1}$\fi}
\def\mo{\ifmmode^{-1}\else$^{-1}$\fi}
\def\sup#1{\ifmmode ^{\rm #1}\else $^{\rm #1}$\fi}
\def\expo#1{\ifmmode \times 10^{#1}\else $\times 10^{#1}$\fi}
\def\,{\thinspace}
\def\lsim{\mathrel{\raise .4ex\hbox{\rlap{$<$}\lower 1.2ex\hbox{$\sim$}}}}
\def\gsim{\mathrel{\raise .4ex\hbox{\rlap{$>$}\lower 1.2ex\hbox{$\sim$}}}}

\def\simprop{\mathrel{\raise .4ex\hbox{\rlap{$\propto$}\lower 1.2ex\hbox{$\sim$}}}}
\def\deg{\ifmmode^\circ\else$^\circ$\fi}
\def\pdeg{\ifmmode $\setbox0=\hbox{$^{\circ}$}\rlap{\hskip.11\wd0 .}$^{\circ}
          \else \setbox0=\hbox{$^{\circ}$}\rlap{\hskip.11\wd0 .}$^{\circ}$\fi}
\def\arcs{\ifmmode {^{\scriptstyle\prime\prime}}
          \else $^{\scriptstyle\prime\prime}$\fi}
\def\arcm{\ifmmode {^{\scriptstyle\prime}}
          \else $^{\scriptstyle\prime}$\fi}
\newdimen\sa  \newdimen\sb
\def\parcs{\sa=.07em \sb=.03em
     \ifmmode \hbox{\rlap{.}}^{\scriptstyle\prime\kern -\sb\prime}\hbox{\kern -\sa}
     \else \rlap{.}$^{\scriptstyle\prime\kern -\sb\prime}$\kern -\sa\fi}
\def\parcm{\sa=.08em \sb=.03em
     \ifmmode \hbox{\rlap{.}\kern\sa}^{\scriptstyle\prime}\hbox{\kern-\sb}
     \else \rlap{.}\kern\sa$^{\scriptstyle\prime}$\kern-\sb\fi}
\def\ra[#1 #2 #3.#4]{#1\sup{h}#2\sup{m}#3\sup{s}\llap.#4}
\def\dec[#1 #2 #3.#4]{#1\deg#2\arcm#3\arcs\llap.#4}
\def\deco[#1 #2 #3]{#1\deg#2\arcm#3\arcs}
\def\rra[#1 #2]{#1\sup{h}#2\sup{m}}

\def\dots{\relax\ifmmode \ldots\else $\ldots$\fi}
\def\WHzsr{\ifmmode $W\,Hz\mo\,sr\mo$\else W\,Hz\mo\,sr\mo\fi}
\def\mHz{\ifmmode $\,mHz$\else \,mHz\fi}
\def\GHz{\ifmmode $\,GHz$\else \,GHz\fi}
\def\mKs{\ifmmode $\,mK\,s$^{1/2}\else \,mK\,s$^{1/2}$\fi}
\def\muKs{\ifmmode \,\mu$K\,s$^{1/2}\else \,$\mu$K\,s$^{1/2}$\fi}
\def\muKRJs{\ifmmode \,\mu$K$_{\rm RJ}$\,s$^{1/2}\else \,$\mu$K$_{\rm RJ}$\,s$^{1/2}$\fi}
\def\muKHz{\ifmmode \,\mu$K\,Hz$^{-1/2}\else \,$\mu$K\,Hz$^{-1/2}$\fi}
\def\MJysr{\ifmmode \,$MJy\,sr\mo$\else \,MJy\,sr\mo\fi}
\def\MJysrmK{\ifmmode \,$MJy\,sr\mo$\,mK$_{\rm CMB}\mo\else \,MJy\,sr\mo\,mK$_{\rm CMB}\mo$\fi}
\def\microns{\ifmmode \,\mu$m$\else \,$\mu$m\fi}

\def\muK{\ifmmode \,\mu$K$\else \,$\mu$\hbox{K}\fi}
\def\microK{\ifmmode \,\mu$K$\else \,$\mu$\hbox{K}\fi}
\def\muW{\ifmmode \,\mu$W$\else \,$\mu$\hbox{W}\fi}
\def\kms{\ifmmode $\,km\,s$^{-1}\else \,km\,s$^{-1}$\fi}
\def\kmsMpc{\ifmmode $\,\kms\,Mpc\mo$\else \,\kms\,Mpc\mo\fi}
\providecommand{\sorthelp}[1]{}

\def\NHUNIT{\ifmmode {\rm \,cm^{-2}} \else $\rm \,cm^{-2}$ \fi} 

\def\wmap{\WMAP}

\def\muKcmb{\ifmmode \,\mu$K$_{\rm CMB}$\else \,$\mu$K$_{\rm CMB}$\fi}
\newcommand{\planck}{\Planck}
\newcommand{\WMAP}{WMAP}

\newcommand{\OmegaM}{\ifmmode\Omega_{\rm M}\else $\Omega_{\rm M}$\fi}

\def\WMAP{{WMAP}}

\newcommand{\commander}{{\tt Commander}}

\providecommand{\Planck}{\textit{Planck}}
\providecommand{\planck}{\Planck}

\providecommand{\text}[1]{\rm{#1}}

\providecommand{\muK}{\mu\rm{K}}

\newcommand{\begm}{\begin{pmatrix}}
\newcommand{\enm}{\end{pmatrix}}

\def\pmb#1{\setbox0=\hbox{#1}%
    \kern-.025em\copy0\kern-\wd0
    \kern.05em\copy0\kern-\wd0
    \kern-.025em\raise.0433em\box0}

\def\p2Y{\;_2Y}
\def\m2Y{\;_{-2}Y}
\def\beglet{
  \addtocounter{equation}{1}%
  \setcounter{parentequation}{\value{equation}}%
  \setcounter{equation}{0}%
  \def\theequation{\arabic{parentequation}\alph{equation}}%
  \ignorespaces
}
\def\endlet{
  \setcounter{equation}{\value{parentequation}}%
  \def\theequation{\arabic{equation}}%
}
\providecommand{\beglet}{\begin{subequations}}
\providecommand{\endlet}{\end{subequations}}

\newcommand{\mksym}[1]{\ifmmode {\rm #1}\else #1\fi}

\providecommand{\text}[1]{\rm{#1}}

\providecommand{\muK}{\mu\rm{K}}

\providecommand{\healpix}{\texttt{HEALPix}}

\newcommand\ba{\begin{eqnarray}}
\newcommand\ea{\end{eqnarray}}
\newcommand\bea{\begin{eqnarray}}
\newcommand\eea{\end{eqnarray}}

\newcommand\be{\begin{equation}}
\newcommand\ee{\end{equation}}

\newcommand{\srolltwo}{{\tt SRoll2}}

\newcommand{\noiseMC}{{\tt S+N}}

\newcommand{\combdata}{\Planck\ LFI$+$WMAP}
\newcommand{\crossdata}{\Planck\  HFI 100$\times$143}

\title{\boldmath Lack-of-correlation anomaly in CMB large scale polarisation maps}

\author[a,b,1]{C.Chiocchetta,\note{Corresponding author.}}
\author[c,d]{A.Gruppuso,}
\author[b]{M.Lattanzi,}
\author[a,b]{P.Natoli,}
\author[a,b]{and L.Pagano}

\affiliation[a]{Dipartimento di Fisica e Scienze della Terra, Universit\'a degli Studi di Ferrara, via Giuseppe Saragat 1, 44122 Ferrara, Italy}
\affiliation[b]{Istituto Nazionale di Fisica Nucleare (INFN), Sezione di Ferrara, Via Giuseppe Saragat 1, 44122 Ferrara, Italy}
\affiliation[c]{INAF - OAS Bologna, Istituto Nazionale di Astrofisica - Osservatorio di Astrofisica e Scienza dello Spazio di Bologna, via Gobetti 101, 40129 Bologna, Italy}
\affiliation[d]{Istituto Nazionale di Fisica Nucleare (INFN), Sezione di Bologna, viale Berti Pichat 6/2, 40127, Bologna, Italy}

\emailAdd{caterina.chiocchetta@unife.it}
\emailAdd{gruppuso@iasfbo.inaf.it}
\emailAdd{lattanzi@fe.infn.it}
\emailAdd{paolo.natoli@unife.it}
\emailAdd{luca.pagano@unife.it}

\abstract{We present an assessment of the CMB large scale anomalies in polarisation using the two-point correlation function as a test case.
We employ the state of the art of large scale polarisation datasets: the first based on a \Planck\ 2018 HFI 100 and 143 GHz cross-spectrum analysis, based on \srolltwo\ processing, and the second from a map-based approach derived through a joint treatment of \Planck\ 2018 LFI and \wmap-9yr. We consider the well-known $S_{1/2}$ estimator, which measures the distance of the two-point correlation function from zero at angular scales larger than $60^{\circ}$, and rely on realistic simulations for both datasets to assess confidence intervals. By focusing on the pure polarisation field described by either the $Q$ and $U$ Stokes parameters or by the local $E-$modes, we show that the first description is heavily influenced by the quadrupole (which is poorly constrained in both datasets) while the second one is more suited for an analysis containing higher multipoles up to $\ell \sim 10$, limit above which both datasets become markedly noise dominated. We find that both datasets exhibit a lack-of-correlation anomaly in pure polarisation, similar to the one observed in temperature, which is better constrained by the less noisy \crossdata\ data, where its significance lies at about $99.5\%$.
We perform our analysis using realizations that are either constrained or non-constrained by the observed temperature field, and find similar results in the two cases.}

\begin{document}
\maketitle
\flushbottom

\section{Introduction}
\label{sec:intro}


The cosmic microwave background (CMB) is one of the most important cosmological observables and has greatly contributed to the success of the standard $\Lambda$CDM model.
Nonetheless, anomalous features exist in the CMB large-angle anisotropy pattern which are in tension with the predictions of $\Lambda$CDM. 
Their statistical significance is assessed at the $2-3\sigma$ level depending on the particular estimator chosen.
Several CMB anomalies exist \cite{Schwarz:2015cma}. In the following we will focus on the lack-of-correlation anomaly, which consists of a suppression in the CMB two-point correlation function at large angular scales with respect to the best-fit $\Lambda$CDM model \cite{Spergel:2003cb, Bernui:2006ft, Copi:2006tu,Copi:2008hw,Copi:2010na}.
This anomaly is directly connected to the so-called CMB lack-of-power anomaly, for which the lack of correlation shows up as a reduction of anisotropy power at large angular scales \cite{Ade:2013zuv,Aghanim:2015xee,Gruppuso:2017nap, Gruppuso:2015xqa, Gruppuso:2015zia}
and to others as well \cite{Monteserin:2007fv, Cruz:2010ud, Gruppuso:2013xba, Muir:2018hjv}.

Two independent experiments, WMAP and \planck\, \cite{Gruppuso:2013dba, Copi:2013cya, Ade:2013nlj, Ade:2015hxq}, agree well on these deviations, therefore limiting (but not completely excluding) the possibility of an instrumental origin. 
An alternative astrophysical explanation of these anomalies is the possible presence of residuals of Galactic emission. 
However, this explanation seems unlikely given that foreground cleaning at large angular scales is usually performed on maps and an imperfect subtraction would normally result into an increase rather than a decrease of 
power\footnote{A possible exception is the presence of chance correlations between foreground and the CMB which appears also unlikely.}. 
A pragmatic approach is therefore to consider the CMB anomalies as correctly measured features in the CMB temperature pattern and assess their statistical significance, e.g. including correctly look-elsewhere effects \cite{Natale:2019dqm}.

If we accept the above point of view, then there are two possible explanations for these features: either we live in a rare (yet not exceedingly rare) realisation of a $\Lambda$CDM cosmology or 
we need a modification of $\Lambda$CDM to account for them.
Discriminating between these two hypotheses can only happen based on some acceptable a posteriori probabilities to exceed. 
Unfortunately, the anomalies show up at large angular scales where the temperature field is already cosmic variance limited, so any additional data, while always useful for consistency tests, are not going to boost statistical significance. 

Improvements can however be expected by including the CMB polarisation pattern, whose measurements are still far from reaching cosmic variance accuracy especially at large scales, where the systematic error budget is currently non negligible\footnote{The power spectrum of \crossdata\ is cosmic variance dominated between $\ell=3$ and $\ell=5$, see figure 10 of \cite{Pagano:2019tci}.}. 
Several, anomalies oriented, analyses that include CMB polarisation have been performed on the \planck\ legacy data \cite{Akrami:2019bkn}, 
using various estimators to quantify statistical significance jointly in temperature and polarization. 
The most adopted estimator has been proposed in \cite{Copi:2013zja} and only uses temperature to E-mode correlations (TE). 
It has been extended to incorporate polarization auto-spectra  EE and BB information in \cite{Yoho:2015bla}. 
The latter is employed, among others, by the \planck\ collaboration in their own analysis \cite{Akrami:2019bkn}. 
Other estimators have been proposed: see for instance \cite{Billi:2019vvg} where a one-dimensional statistic involving TT, TE, EE angular power spectra is employed. 

Incorporating polarization into a joint estimator calls for a choice. 
On the one hand, it is desirable to test whether the polarization observations are consistent with $\Lambda$CDM once the temperature observations are given. 
This can be accomplished by using constrained realizations of the joint temperature and polarization fields.
On the other hand, it is also useful to test the significance of anomalies in temperature and polarization leaving both fields free to fluctuate within the $\Lambda$CDM predictions. 
This implies dealing with unconstrained (i.e.\ open) realisations. For instance, authors in \cite{Copi:2013zja} work under the first assumption, while the \planck\ collaboration \cite{Akrami:2019bkn} assumes the second.

In this paper, we analyse the consequences of either assumption. We employ two datasets:
the first one is based on the cross-spectra between \planck\ 100 and 143 GHz channels obtained with the \srolltwo\ processing \cite{Delouis:2019bub} while,
the second is based on the auto-spectra obtained combining the \Planck\ 70 GHz channel with the Ka, Q and V bands of \wmap\ \cite{Natale:2020owc}.  
These two datasets cannot be easily further combined because a proper combination should happen at map level (as it has been done for \Planck\ 70 GHz and WMAP) but 
the \Planck\ 100 and 143 GHz channels do not allow for this. Therefore we analyse the two datasets separately.
We stress that these datasets have never been employed in the context of CMB anomalies before. This is an aspect where our analysis is entirely novel.

We focus on the lack of correlation anomaly in polarization by considering the correlation functions $C^{QQ}$ and $C^{UU}$, $Q$ and $U$ being the linear polarization Stokes parameters. 
We consider also the $C^{EE}$ correlation function as proposed in \cite{Yoho:2015bla}.
The paper is organised as follows. In Section \ref{sec:dataset} we describe the datasets considered as well as our power spectra estimation procedure employed to derive correlation functions.
In Section \ref{Analysis} we present the estimators used to asses the statistical significance of the considered anomaly.
In Section \ref{sec:results} we set forth our main findings while in Section \ref{sec:conclusions} we draw our conclusions. 

\section{Datasets and methodology}\label{sec:dataset}

We consider the most constraining large-scale polarization datasets currently available, i.e. the cross-spectra between \planck\ 100 and 143 GHz channels \cite{Aghanim:2018fcm} as presented in \cite{Delouis:2019bub,Pagano:2019tci} (hereafter \crossdata) and the auto-spectra obtained combining the \Planck\ 70 GHz channel \cite{Akrami:2018jnw} with the Ka, Q and V bands of \wmap\ \cite{Bennett:2012zja} as presented in \cite{Natale:2020owc} (hereafter \combdata).
Here we briefly provide some general information useful to understand the procedure followed in preparing the former datasets. All the maps contained in the two datasets are mitigated from polarized Galactic foreground emissions (thermal dust and synchrotron) through a template fitting procedure, see e.g., \cite{Page:2006hz,Aghanim:2019ame}. In the \combdata\ dataset the auto-spectra are computed from the CMB map, built through an optimal weighting of the four foreground reduced input maps (i.e. 70 GHz, Ka, Q and V). In temperature both datasets employ the \commander\ \planck\ 2018 CMB solution smoothed trough a Gaussian kernel with FWHM of $440$ arcminutes and downgraded to a \healpix\ \footnote{\url{https://healpix.sourceforge.io/}} ${\rm N}_{\rm side}=16$ resolution \cite{Gorski:2004by}. The polarization maps are instead smoothed assuming a cosine window profile as suggested in \cite{Benabed:2009af,Aghanim:2016yuo}, and re-pixelized to the same \healpix\ resolution as temperature. We select a useful sky fraction of 50\% of  \crossdata\ and 54\% for \combdata\ as suggested respectively in \cite{Pagano:2019tci} and in \cite{Natale:2020owc}.

In order to estimate the angular power spectra from the CMB maps we employ a Quadratic Maximum Likelihood (hereafter QML) method as presented in \cite{Tegmark:2001zv,Gruppuso:2009ab,Aghanim:2016yuo}.
For a given map $\textbf{x} = (\textbf{T, Q, U})$ the QML provides the estimated auto angular power spectra as
\begin{equation}
    \hat{C}^{X}_{\ell}= \sum_{\ell', X'}(F^{-1})_{\ell\ell'}^{X, X'}\left[\textbf{x}^{t}\textbf{E}^{\ell'}_{X'}\textbf{x} -{\rm Tr}(\textbf{NE}^{\ell'}_{X'})\right],
    \label{eq:QMLspectrum}
\end{equation}
where $X$ and $X'$ are one of $TT$, $EE$, $BB$, $TE$, $TB$, $EB$ and $F_{X, X'}^{\ell\ell'}$ is the Fisher information matrix defined as
\begin{equation}
F_{X, X'}^{\ell\ell'}=\frac12{\rm Tr}\left[\textbf{C}^{-1}\frac{\partial \textbf{S}}{\partial C_{\ell}^{X}}\textbf{C}^{-1}\frac{\partial \textbf{S}}{\partial C_{\ell}^{X'}}\right],
    \label{eq:QMLfisher}
\end{equation}
with $\textbf{C} \equiv \textbf{S}( C_{\ell})+\textbf{N}$ being the CMB signal ($\textbf{S}$) plus noise ($\textbf{N}$) covariance matrix and $C_{\ell}$ a fiducial set of CMB angular power spectra.
Finally, the $\textbf{E}$ matrix in eq.\ (\ref{eq:QMLspectrum}) is given by 
\begin{equation}
    \textbf{E}^{\ell}_{X} = \frac{1}{2}\textbf{C}^{-1}\frac{\partial \textbf{S}}{\partial C_{\ell}^{X}}\textbf{C}^{-1}.
    \label{eq:QMLE}
\end{equation}

Assuming uncorrelated noise between two maps $\textbf{x}_a$ and $\textbf{x}_b$, eq.\ (\ref{eq:QMLspectrum}) can be easily extended to cross-spectrum estimation which reads 

\begin{equation}
    \hat{C}^{X}_{\ell}= \sum_{\ell', X'}(F^{-1})_{\ell\ell'}^{X, X'}\textbf{x}_a^{t}\textbf{E}^{\ell'}_{X'}\textbf{x}_b, 
    \label{eq:QMLspectrumcross}
\end{equation}
having coherently defined
\begin{eqnarray}
& &F_{X, X'}^{\ell\ell'}=\frac12{\rm Tr}\left[\textbf{C}_a^{-1}\frac{\partial \textbf{S}}{\partial C_{\ell}^{X}}\textbf{C}_b^{-1}\frac{\partial \textbf{S}}{\partial C_{\ell}^{X'}}\right],
    \label{eq:QMLfishercross}\\
        & &\textbf{E}^{\ell}_{X} = \frac{1}{2}\textbf{C}_a^{-1}\frac{\partial \textbf{S}}{\partial C_{\ell}^{X}}\textbf{C}_b^{-1},
    \label{eq:QMLE}\\
    & &\textbf{C}_a = \textbf{S}( C_{\ell})+\textbf{N}_a,\nonumber\\
   & &\textbf{C}_b = \textbf{S}( C_{\ell})+\textbf{N}_b.\nonumber
\end{eqnarray}

For the two aforementioned datasets we also consider a set of $500$ noise plus residual systematics simulations described in the two dedicated papers \cite{Delouis:2019bub,Pagano:2019tci}. 

\section{Analysis}\label{Analysis}

\begin{figure}[tbp]
\centering 
\includegraphics[width=.45\textwidth]{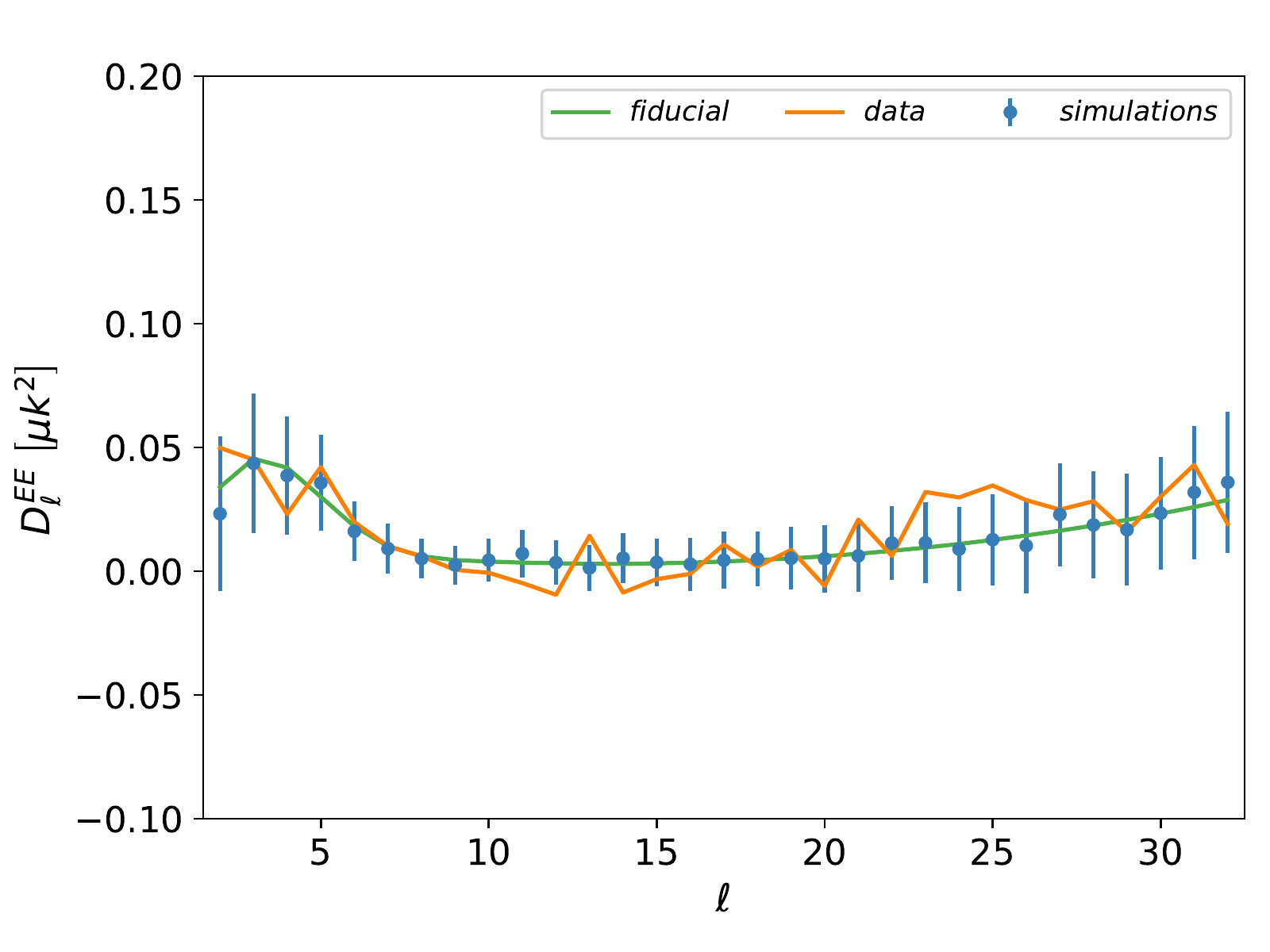}
\hfill
\includegraphics[width=.45\textwidth]{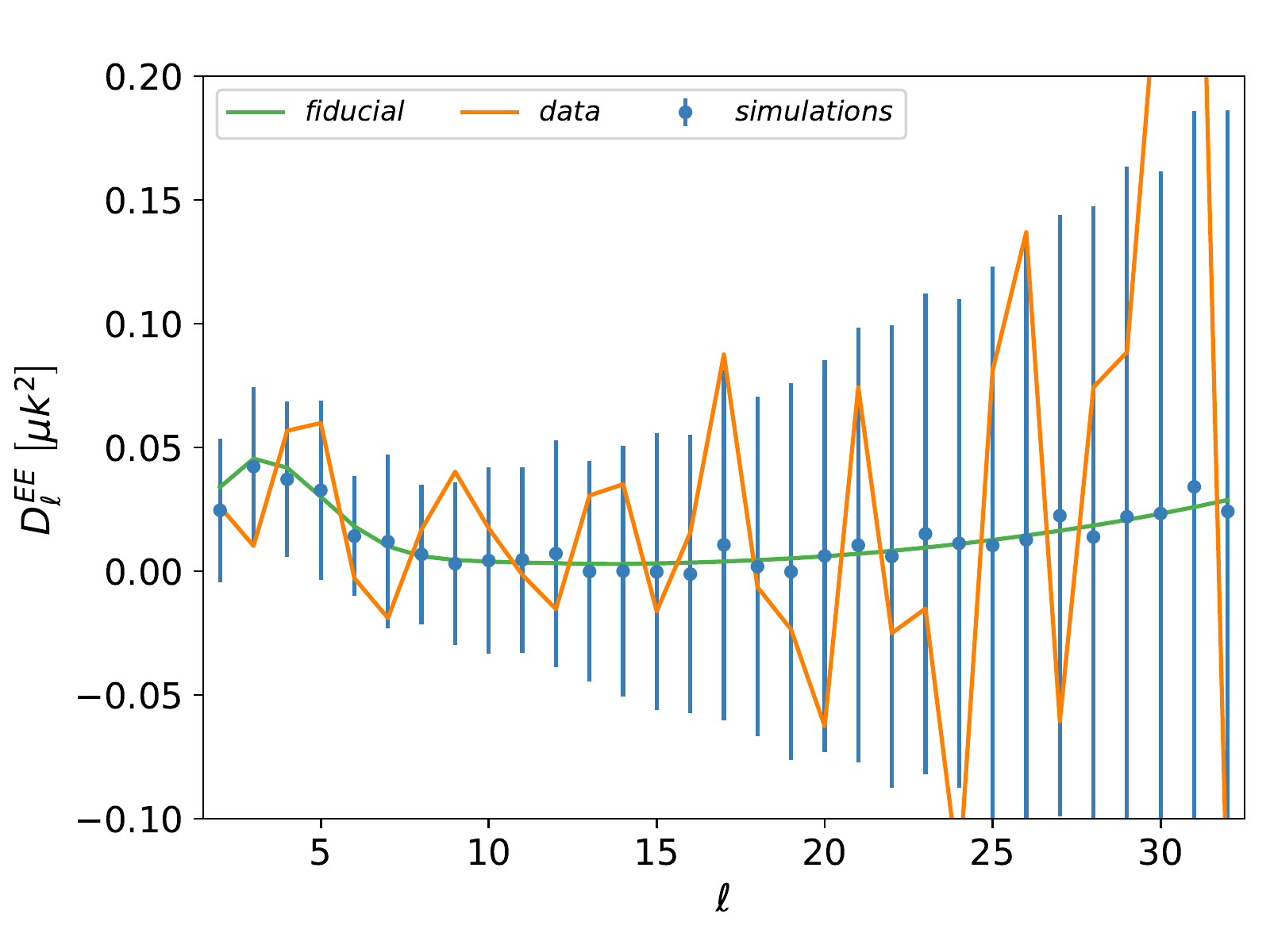}
\caption{\label{fig:postqml_ps} E-mode angular power spectrum $D_{\ell}^{EE} \equiv \ell(\ell+1) C_{\ell}^{EE}/2\pi$ for \crossdata\ GHz (left panel) and \combdata\ (right panel). The orange line is the data power spectrum while blue dots are the mean of power spectra extracted from 250000 constrained maps. The error bars are computed as the standard deviation of the simulations. Note that the range of values on the y-axis is the same for both panels.}
\end{figure}

As discussed in the introduction we need both constrained and unconstrained simulations of the temperature and polarization fields.
To build the set of constrained realizations we follow the procedure described in \cite{Copi:2013zja} and generate the polarized spherical harmonic coefficients, $a^{E}_{\ell m}$ and $a^{B}_{\ell m}$, as
 \begin{eqnarray}
    a^{E}_{\ell m} &=& \frac{C_{\ell}^{TE}}{C_{\ell}^{TT}}a^{T_{data}}_{\ell m} + \zeta_{1}\sqrt{C_{\ell}^{EE} - \frac{(C_{\ell}^{TE})^{2}}{C_{\ell}^{TT}}},\nonumber\\
 a^{B}_{\ell m} &=&\zeta_{2}\sqrt{C_{\ell}^{BB}},
 \label{almconst}
\end{eqnarray}
where $\zeta_{1}$ and $\zeta_{2}$ are random Gaussian realizations with zero mean and unit variance, $C_{\ell}^{XX}$ are the spectra corresponding to the \planck\ best fit $\Lambda$CDM model and $a^{T_{data}}_{\ell m}$ are extracted from the observed temperature map. When building maps from spherical harmonic coefficients, we apply the same aforementioned window functions. Finally, we combine each of the 500 simulated CMB signal maps with all the 500 noise maps described in the previous section, forming a set of 250000 signal plus noise (hereafter \noiseMC) realizations.
In figure \ref{fig:postqml_ps} we show the E-mode power spectra of the data maps compared with mean and standard deviations of the corresponding spectra of our \noiseMC\ Monte Carlo. Both \crossdata\ and \combdata\ spectra do not show any evident outlier when compared to \noiseMC\ simulations. 

We start computing the signal-to-noise ratio, $S/N$,
\begin{equation} 
    \frac{S}{N} = \sqrt{\sum_{\ell=2}^{\ell_{\rm max}} \left(\frac{C^{EE}_{\ell}}{\sigma_{\ell}}\right)^{2}},
\end{equation}
where $C^{EE}_{\ell}$ is a fiducial power spectrum, $\sigma_{\ell}$ is the standard deviation of the \noiseMC\ simulations and $\ell_{\rm max}$ is the maximum multipole considered in the sum.
\begin{figure}[tbp]
\centering
\includegraphics[scale = 0.50]{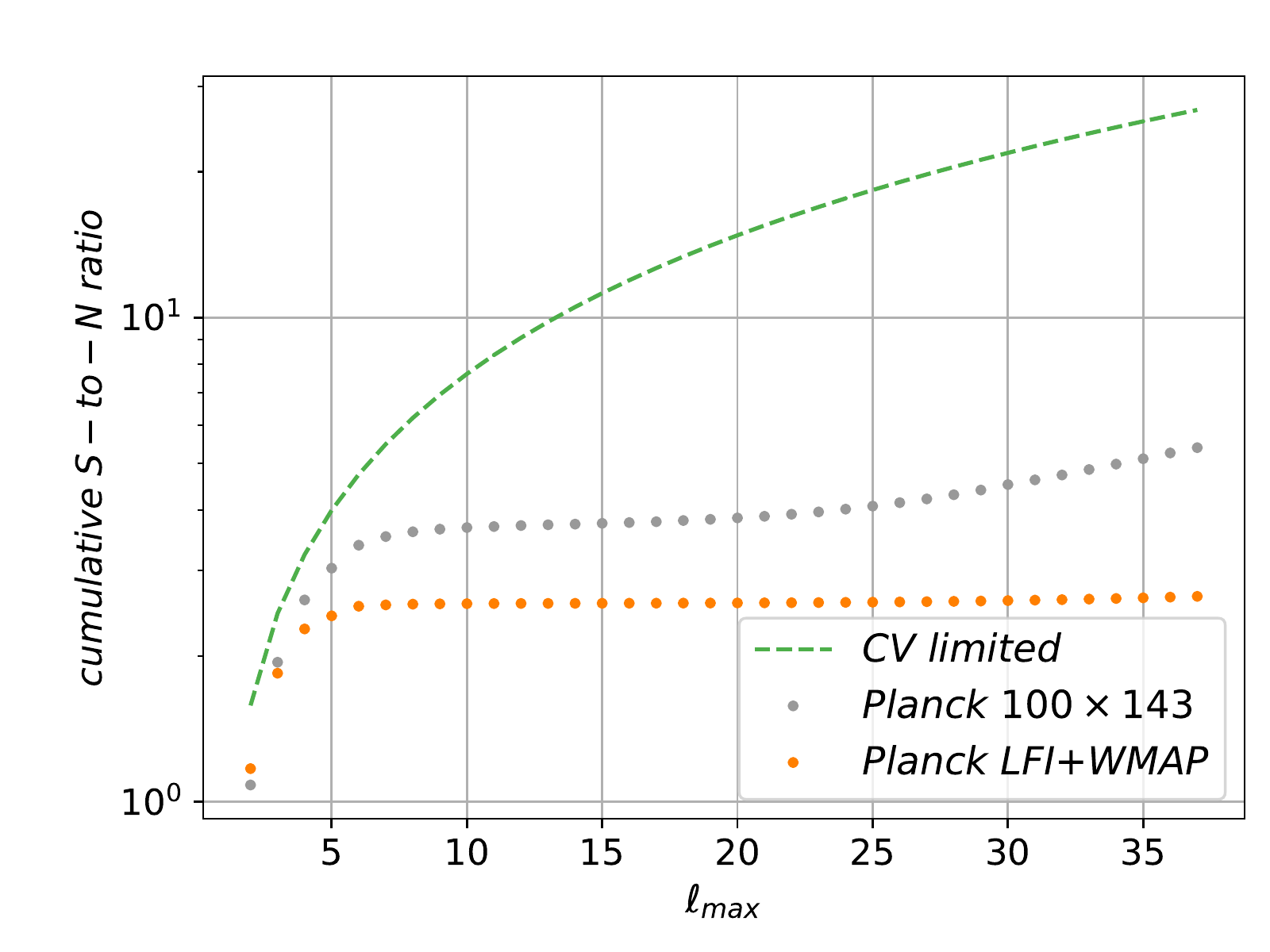}
\caption{\label{fig:rapp_var}Comparison of the integrated signal to noise ratio for the \crossdata\ (grey dots) and \combdata\ (orange dots) dataset and for a sample of ideal simulations (green dashed line). Both datasets show a plateau above $\ell_{\rm max} \simeq10$, which is consequently chosen as maximum multipole in the analysis. For \crossdata\ a rise in trend is visible at $\ell_{\rm max} >20$, due to the corresponding increase in the E-mode signal. This effect can be safely ignored in our analysis.}
\end{figure}
In Figure \ref{fig:rapp_var} we show $S/N$ as a function of $\ell_{\rm max}$ for our two datasets and for a cosmic variance limited full sky survey. 
As expected, \crossdata\ has a better $S/N$ ratio with respect to \combdata\ for almost all the $\ell_{\rm max}$ considered. The only exception is the quadrupole, where the \crossdata\ variance is dominated by residual dipole leakage (see \cite{Delouis:2019bub,Pagano:2019tci} for details). 
Both datasets considered show a plateau above $\ell_{\rm max} \simeq10$ where the variance of the noise starts dominating the total variance.
This justifies our choice of  $\ell_{\rm max}=10$ as  maximum multipole in the following analysis.

\subsection{Estimators}\label{subsec:estimators}

We focus on an estimator originally suggested by the WMAP team, called $S_{1/2}$ \cite{Spergel:2003cb}. 
The idea is to measure the distance between the correlation function and zero over a chosen range of angles \cite{Copi:2008hw}.
We formally define this estimator below.

\subsubsection{Temperature}\label{subsubsec:temperature}
We start by reviewing the definition of $S_{1/2}$ in temperature,
whose fluctuations are usually expanded in terms of scalar spherical harmonics: 
\begin{equation}
    \frac{\Delta T(\hat{n})}{T_{0}} \equiv \Theta(\hat{n}) = \sum_{\ell m}  a_{\ell m}^{T}Y_{\ell m}(\hat{n}) \, ,
    \label{temp_perturbations}
\end{equation}
and the covariance of the coefficients $a_{\ell m}^T$ defines the anisotropy angular power spectrum,
\begin{equation}
    \langle a_{\ell \textit{m}}^T a_{\ell' \textit{m}'}^{T \star}\rangle = C^{TT}_{\ell}\delta_{\ell \ell'}\delta_{\textit{m}\textit{m}'} \, ,
    \label{power_spectrum}
\end{equation}
standing the assumption of statistical isotropy and independence of the modes.
The angular power spectrum, $C^{TT}_{\ell}$, and the two-point angular correlation function, $C(\theta)$, are related by the following expression,
\begin{equation}
    C(\theta) \equiv \langle \Theta\left(\hat n_{1}\right)\Theta\left(\hat n_{2}\right)\rangle = \frac{1}{4\pi}\sum_{\ell=0}^{\infty}\left(2\ell+1\right)C^{TT}_{\ell}\mathcal{P}_{\ell}\left(\cos(\theta)\right),
    \label{ctheta_pws}
\end{equation}
where $\hat n_{1} \cdot \hat n_{2} = \cos(\theta)$ and $\mathcal{P}_{\ell}$ are the Legendre Polynomials.
The $S_{1/2}$ statistic in temperature, is defined as
\begin{equation}
    S^{TT}_{1/2}\equiv \int_{-1}^{1/2}d(\cos{\theta})[C^{TT}(\theta)]^{2} \, ,
    \label{EstimatorTT}
\end{equation}
and is used to quantify the lack of correlation at scales larger than $60^{\circ}$.
Substituting \eqref{ctheta_pws} into \eqref{EstimatorTT} we can rewrite the estimator in terms of the angular power spectrum,
\begin{equation}
    S^{TT}_{1/2} = \sum_{\ell = 2}^{\ell_{max}} \sum_{\ell' = 2}^{\ell'_{max}}\frac{(2\ell+1)}{4\pi}\frac{(2\ell'+1)}{4\pi}C_{\ell}^{TT}I_{\ell\ell'}C_{\ell'}^{TT},
    \label{EstimatorTT1}
\end{equation}
where the matrix $I_{\ell\ell'}$ is defined as 
\begin{equation}
    I_{\ell \ell'}(x) \equiv \int_{-1}^{x} \mathcal{P}_{\ell}(x')\mathcal{P}_{\ell'}(x')dx',
    \label{matrixTT}
\end{equation}
and evaluated at $x=1/2$, with $x = \cos{\theta}$.

\subsubsection{Polarization}
Linear polarization is a spin-$2$ quantity and can be described by the Stokes parameters $Q$ and $U$ \cite{Kamionkowski:1996ks}. In analogy with T we can define the corresponding two-point angular correlation function as $C^{QQ}(\theta) = \langle Q_{r}(\hat{n}_{1})Q_{r}(\hat{n}_{2})\rangle$ and $C^{UU}(\theta) = \langle U_{r}(\hat{n}_{1})U_{r}(\hat{n}_{2})\rangle$. The Stokes parameters appearing in the correlation functions are defined with respect to a reference frame on the tangent plane with axes parallel and perpendicular to the great arch connecting $\hat n_1$ and $\hat n_2$. 
As in \cite{Kamionkowski:1996ks} we choose one point to be the north pole and the other on $\phi = 0$ longitude. This choice is denoted by the suffix $r$ in the above definitions of the correlations functions.
The coordinate system is hence fixed and the correlation functions depend only on the separation $\theta$ between $\hat n_1$ and $\hat n_2$.
The definition of $S_{1/2}$ in polarisation is analogous to the temperature case,
\begin{equation}
    S^{QQ,UU}_{1/2}\equiv \int_{-1}^{1/2}d(\cos{\theta})[C^{QQ,UU}(\theta)]^{2},
    \label{estimator_pol}
\end{equation}
but it is useful again to rewrite it in terms of the angular power spectrum. 
The Stokes parameters can be decomposed using spin-2 spherical harmonics:
\begin{equation}
    Q(\hat{n})\pm iU(\hat{n}) = \sum_{\ell, m}  {}_{\pm 2}a_{\ell, m}^{P}{}_{\pm 2}Y_{\ell,m}(\hat{n}).
    \label{pol_decomposition}
\end{equation}
A linear combination of the spin-2 spherical harmonic coefficients $_{\pm 2}a_{\ell, m}^{P}$ gives the E- and B-mode coefficients, $a^{E/B}_{\ell,m}$, which are two scalar quantities \cite{Kamionkowski:1996ks}:
\begin{subequations}
    \label{pol_alm}
    \begin{align}
    a^{E}_{\ell,m} &= -\frac{1}{2}[{}_{2}a^{P}_{\ell,m} + {}_{-2}a^{P}_{\ell,m}], \label{sub_pol_alm1} \\
    a^{B}_{\ell,m} &= + \frac{i}{2}[{}_{2}a^{P}_{\ell,m} - {}_{-2}a^{P}_{\ell,m}].
    \label{sub_pol_alm2}
    \end{align}
\end{subequations}
The E- and B-mode power spectra, $C^{EE/BB}_{\ell}$, are defined as:
\begin{subequations}
    \label{pol_pws}
    \begin{align}
    \langle a^{E}_{\ell \textit{m}}a_{\ell' \textit{m}'}^{E\star}\rangle &= C^{EE}_{\ell}\delta_{\ell \ell'}\delta_{\textit{m}\textit{m}'},\\
    \langle a^{B}_{\ell \textit{m}}a_{\ell' \textit{m}'}^{B\star}\rangle &= C^{BB}_{\ell}\delta_{\ell \ell'}\delta_{\textit{m}\textit{m}'}.
    \label{sub_pol_pws}
    \end{align}
\end{subequations}
We can now express the correlation functions in terms of the power angular spectra \cite{Kamionkowski:1996ks},
\begin{subequations}
    \label{eq:corrfuncQU_pws}
    \begin{align}
    C^{QQ}(\theta) &= - \sum_{\ell} \frac{2\ell+1}{4\pi}\left(\frac{2(\ell-2)!}{(\ell+2)!}\right)[C_{\ell}^{EE}G^{+}_{\ell 2}(\cos(\theta))+C_{\ell}^{BB}G^{-}_{\ell 2}(\cos(\theta))],\\
    C^{UU}(\theta) &= - \sum_{\ell} \frac{2\ell+1}{4\pi}\left(\frac{2(\ell-2)!}{(\ell+2)!}\right)[C_{\ell}^{BB}G^{+}_{\ell 2}(\cos(\theta))+C_{\ell}^{EE}G^{-}_{\ell 2}(\cos(\theta))],
    \label{sub_corrfuncQU_pws}
    \end{align}
\end{subequations}
where
\begin{subequations}
    \label{Gfunc}
    \begin{align}
    G^{+}_{\ell m}(\cos\theta) &= -\left(\frac{\ell-m^{2}}{\sin^{2}{\theta}} + \frac{\ell(\ell+1)}{2}\right)\mathcal{P}_{\ell}^{m}(\cos{\theta})+(\ell+m)\frac{\cos{\theta}}{\sin^{2}{\theta}}\mathcal{P}_{l-1}^{m}(\cos{\theta})\label{sub_Gp},\\
    G^{-}_{\ell m}(\cos\theta) &= \frac{m}{sin^{2}\theta}\left((\ell-1)\cos(\theta)\mathcal{P}_{\ell}^{m}(\cos{\theta}) - (\ell+m)\mathcal{P}_{l-1}^{m}(\cos{\theta}) \right)
    \label{sub_Gm},
    \end{align}
\end{subequations}
being the $\mathcal{P}^{m}_{\ell}(\cos{\theta})$ the associated Legendre polynomials.
Plugging eq. \eqref{eq:corrfuncQU_pws} into \eqref{estimator_pol} we obtain the following expression:
\begin{multline}
    S^{QQ}_{1/2}=  \sum_{\ell=2}^{\ell_{max}}  \sum_{\ell'=2}^{\ell'_{max}}  \frac{2\ell+1}{8\pi} \frac{2\ell'+1}{8\pi}\\
    \left( C^{EE}_{\ell}I^{(1)}_{\ell\ell'}C^{EE}_{\ell'}+C^{BB}_{\ell}I^{(3)}_{\ell\ell'}C^{BB}_{\ell'}+C^{EE}_{\ell}I^{(2)}_{\ell\ell'}C^{BB}_{\ell'}+C^{EE}_{\ell}I^{(4)}_{\ell\ell'}C^{BB}_{\ell'}\right),
    \label{eq:shalfQQ}
\end{multline}
where we have followed the notation of \cite{Yoho:2015bla}. For $S^{UU}_{1/2}$ matrices $I^{(3)}_{\ell\ell'}$ and $I^{(1)}_{\ell\ell'}$ are swapped. More details on this calculation as well as the definition of the $I^{(X)}_{\ell\ell'}$ 
matrices are given in Appendix \ref{a1}.

\subsubsection{Two-point correlation functions for E- and B-modes}
An alternative to Q and U is to express the polarization in terms of local E- and B-modes. These scalar quantities can be obtained from the Stokes parameters through a lowering-spin operator and their spherical harmonic coefficients are defined in eqs. (\ref{sub_pol_alm1}) and (\ref{sub_pol_alm2}).
Their use has been suggested by \cite{Yoho:2015bla} in the context of polarization correlation functions to complement the information given by Q and U, as we will show in the next section. 
The local correlation functions for E and B are defined as:
\begin{subequations}
    \label{EB_pws}
    \begin{align}
    C^{\hat{E}\hat{E}}(\theta) = \langle\hat{E}(\hat{n}_{1})\hat{E}(\hat{n}_{2})\rangle,\\
    C^{\hat{B}\hat{B}}(\theta) = \langle\hat{B}(\hat{n}_{1})\hat{B}(\hat{n}_{2})\rangle,
    \label{sub_EB_pws}
    \end{align}
\end{subequations}
where the $\hat{B}(\hat{n})$ and $\hat{E}(\hat{n})$ functions are expanded as:
\begin{subequations}
    \label{EB_decomposition}
    \begin{align}
    \hat{E}(\hat{n}) = \sum_{\ell,m} \sqrt{\frac{(\ell+2)!}{(\ell-2)!}}a_{\ell,m}^{E}Y_{\ell,m}(\hat{n}),\\
    \hat{B}(\hat{n}) = \sum_{\ell,m} \sqrt{\frac{(\ell+2)!}{(\ell-2)!}}a_{\ell,m}^{B}Y_{\ell,m}(\hat{n}).
    \label{sub_EB_decomposition}
    \end{align}
\end{subequations}
The two-point angular correlation functions can be written in terms of the angular power spectrum, in analogy with temperature:
\begin{subequations}
    \label{eq:EB_local}
    \begin{align}
    &C^{\hat{E}\hat{E}}(\theta) = \sum_{\ell} \frac{2\ell+1}{4\pi}\frac{(\ell+2)!}{(\ell-2)!}C^{EE}_{\ell}\mathcal{P}_{\ell}(\cos{\theta}),\\
    &C^{\hat{B}\hat{B}}(\theta) = \sum_{\ell}\frac{2\ell+1}{4\pi} \frac{(\ell+2)!}{(\ell-2)!}C^{BB}_{\ell}\mathcal{P}_{\ell}(\cos{\theta}),
    \label{sub_EB_local}
    \end{align}
\end{subequations}
and the expressions for the estimators $S^{EE}_{1/2}$ and $S^{BB}_{1/2}$ are:
\begin{equation}
    S^{XX}_{1/2} = \sum_{\ell = 2}^{\ell_{max}} \sum_{\ell' = 2}^{\ell'_{max}}\frac{2\ell+1}{4\pi} \frac{(\ell+2)!}{(\ell-2)!}\frac{2\ell'+1}{4\pi} \frac{(\ell'+2)!}{(\ell'-2)!}C^{XX}_{\ell}I_{\ell,\ell'}C^{XX}_{\ell'},
    \label{eq:shalfXX}
\end{equation}
where $X$ can be $E$ or $B$ and $I_{\ell \ell'}$ is the same kernel defined in eq.\ \eqref{matrixTT} above.

\section{Results}\label{sec:results}

In this section we present results for the correlation functions and for the distribution of the $S_{1/2}$ estimators, for the Q, U and local E-modes fields. For the sake of brevity, we only show plots for constrained simulations, while in Table \ref{tab:results} we report the results for both the constrained and unconstrained case. We start by discussing our results on the Q, U and local E-modes correlation functions, which are needed to better highlight the specificity of the $S^{QQ}_{1/2}$, $S^{UU}_{1/2}$ and $S^{EE}_{1/2}$ estimators.

\begin{figure}[tbp]
\centering 
\includegraphics[width=.45\textwidth]{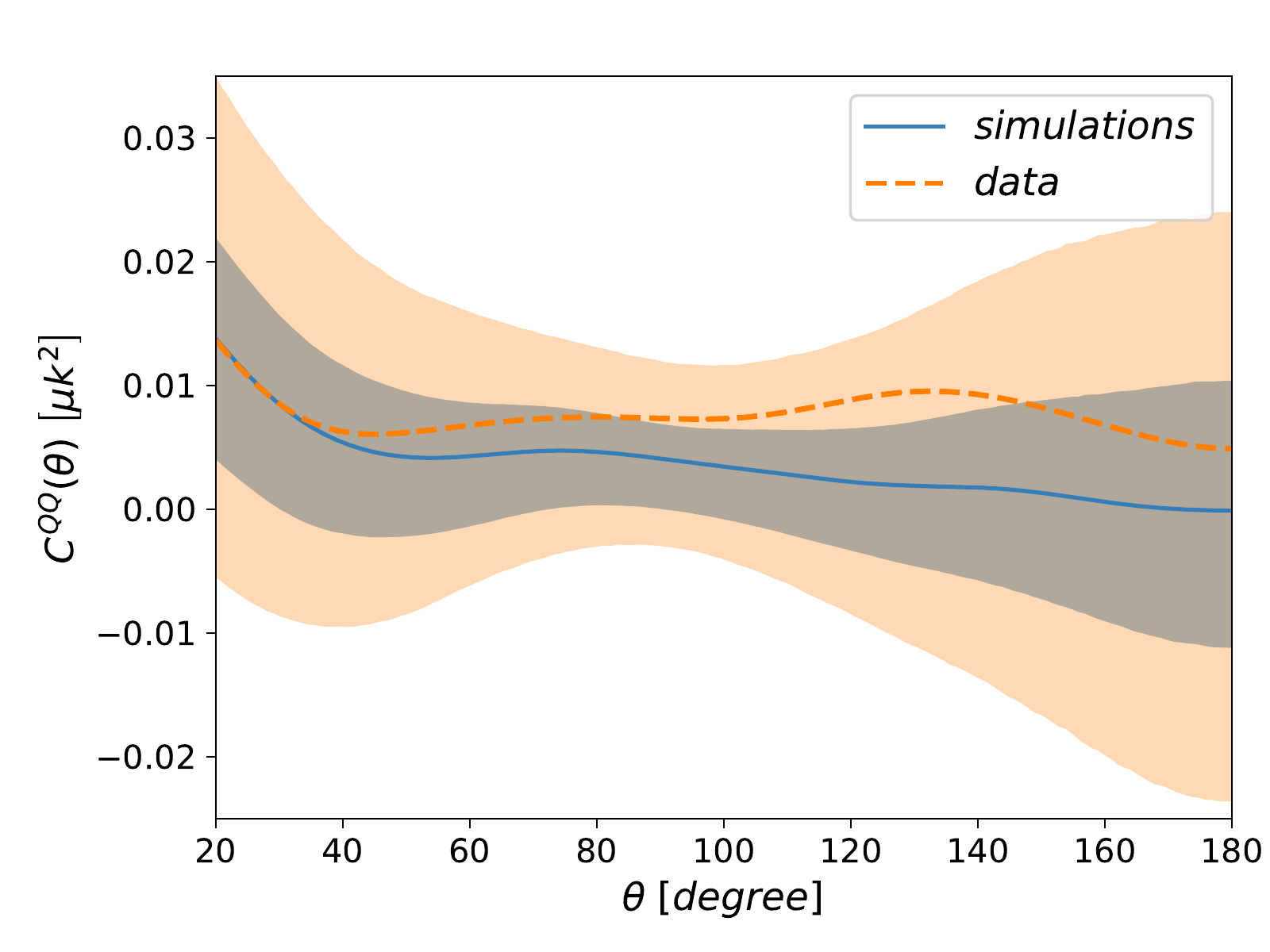}
\hfill
\includegraphics[width=.45\textwidth]{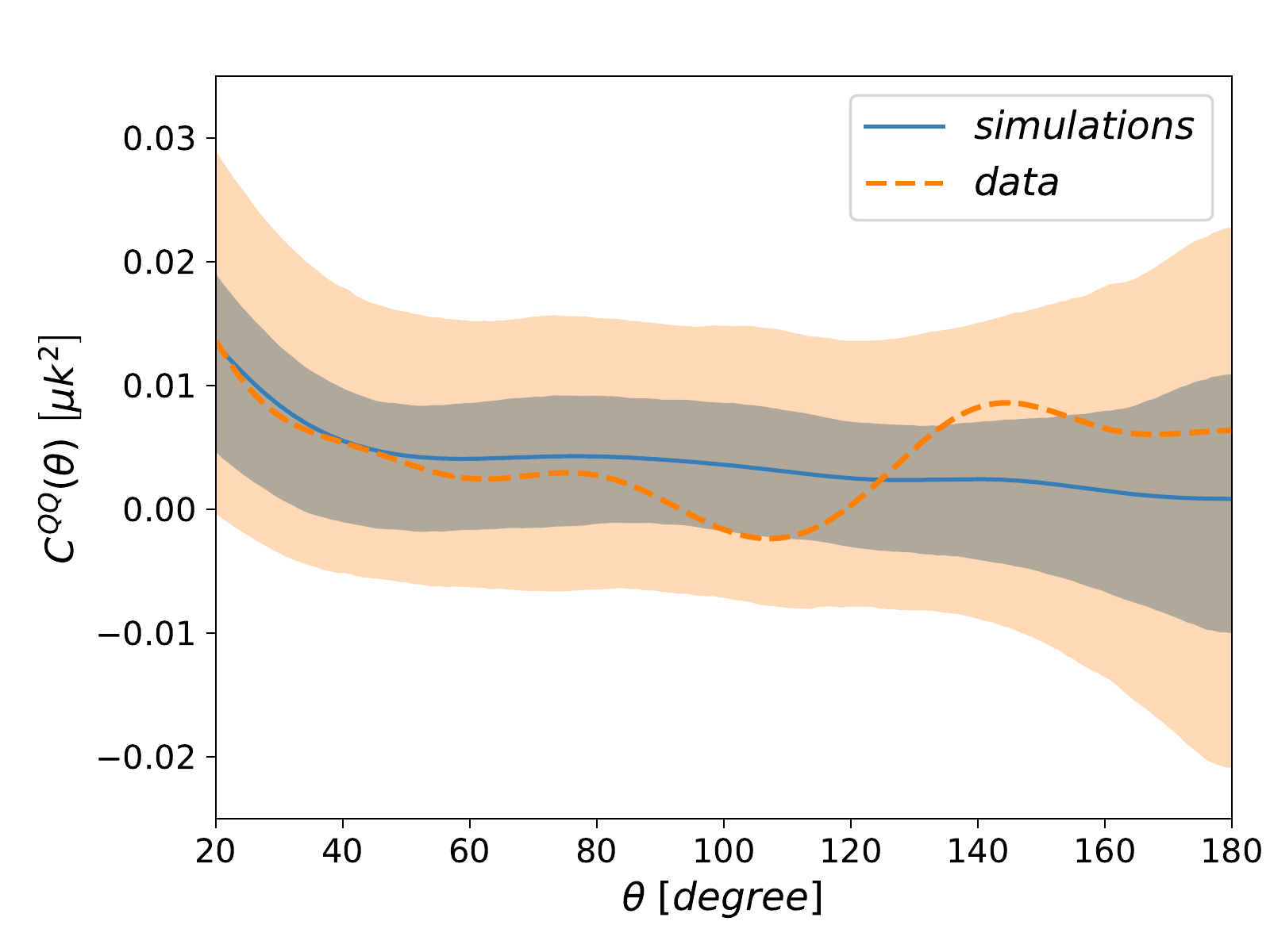}
\caption{\label{fig:cthetaQQ} Two point angular $QQ$ correlation function. The orange dashed line represents data while the blue line is the mean of 250000 constrained simulations. The shaded region represents the $68\%$ and $95\%$ C.L.. The datasets employed are \crossdata\ on the left and \combdata\ on the right. The analysis is performed up to $\ell=10$}
\end{figure}

\begin{figure}[tbp]
\centering 
\includegraphics[width=.45\textwidth]{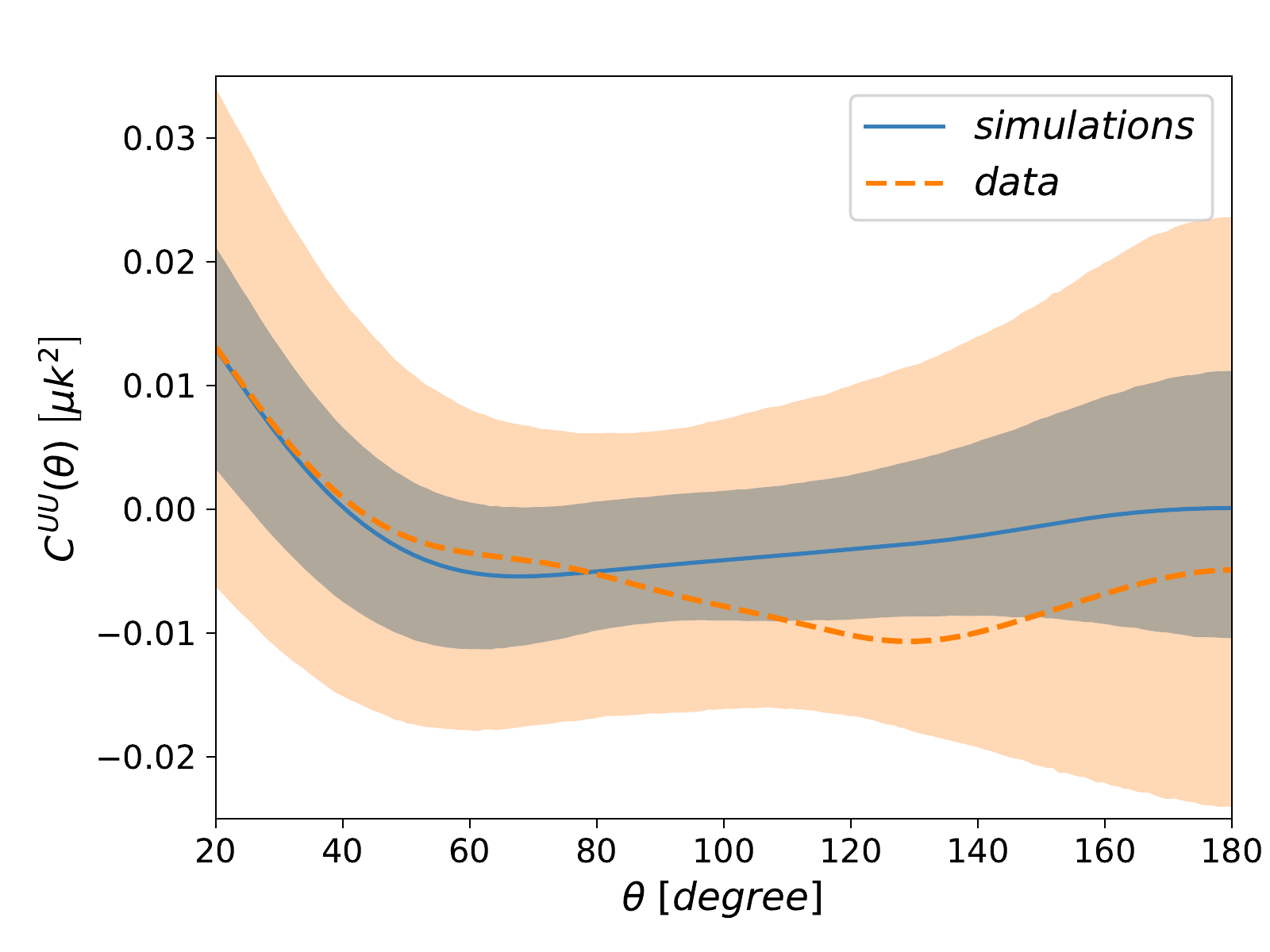}
\hfill
\includegraphics[width=.45\textwidth]{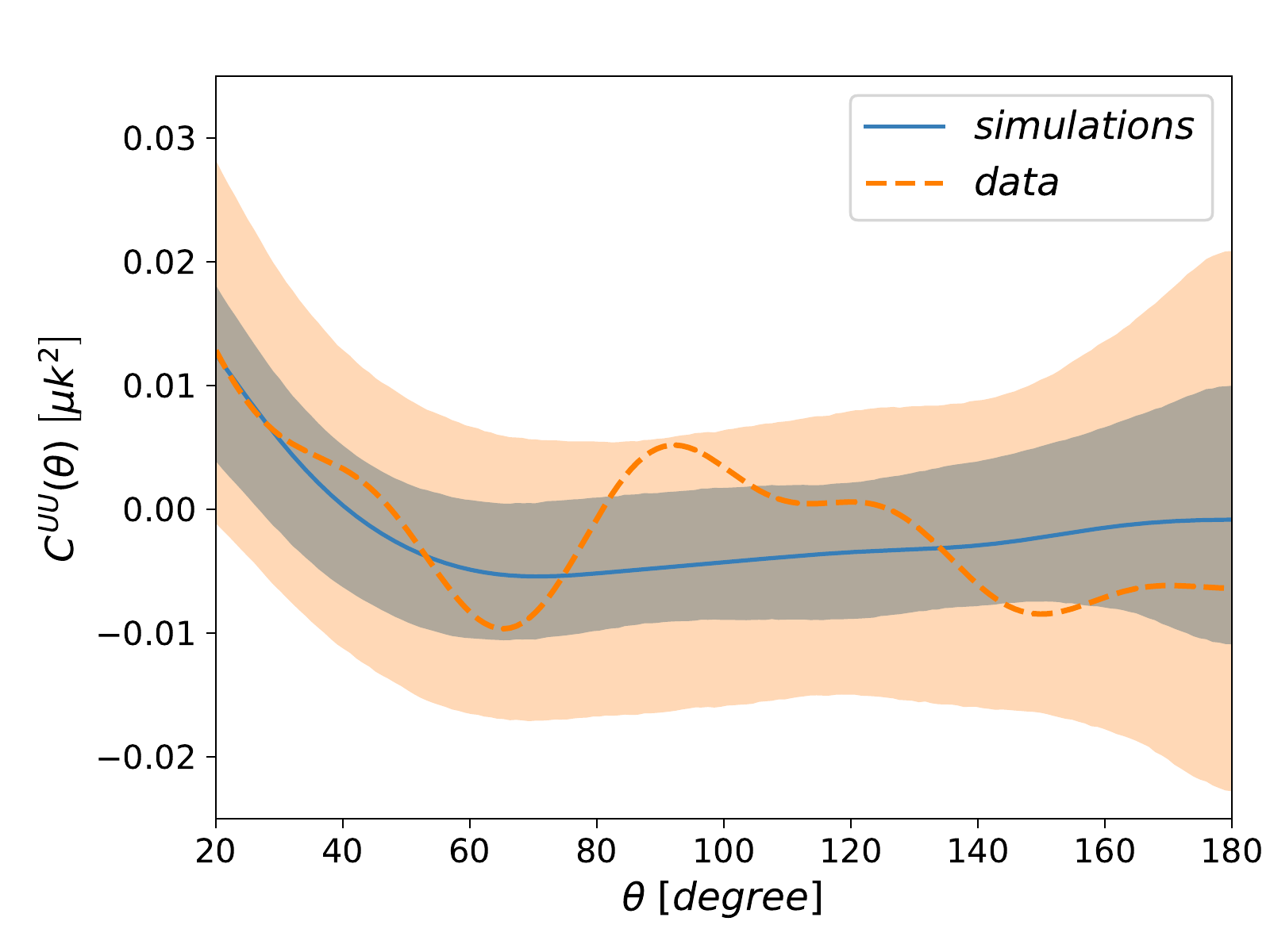}
\caption{\label{fig:cthetaUU}  As in Fig.~\ref{fig:cthetaQQ} above but for the $UU$ correlation function instead of $QQ$.}
\end{figure}

\begin{figure}[tbp]
\centering 
\includegraphics[width=.45\textwidth]{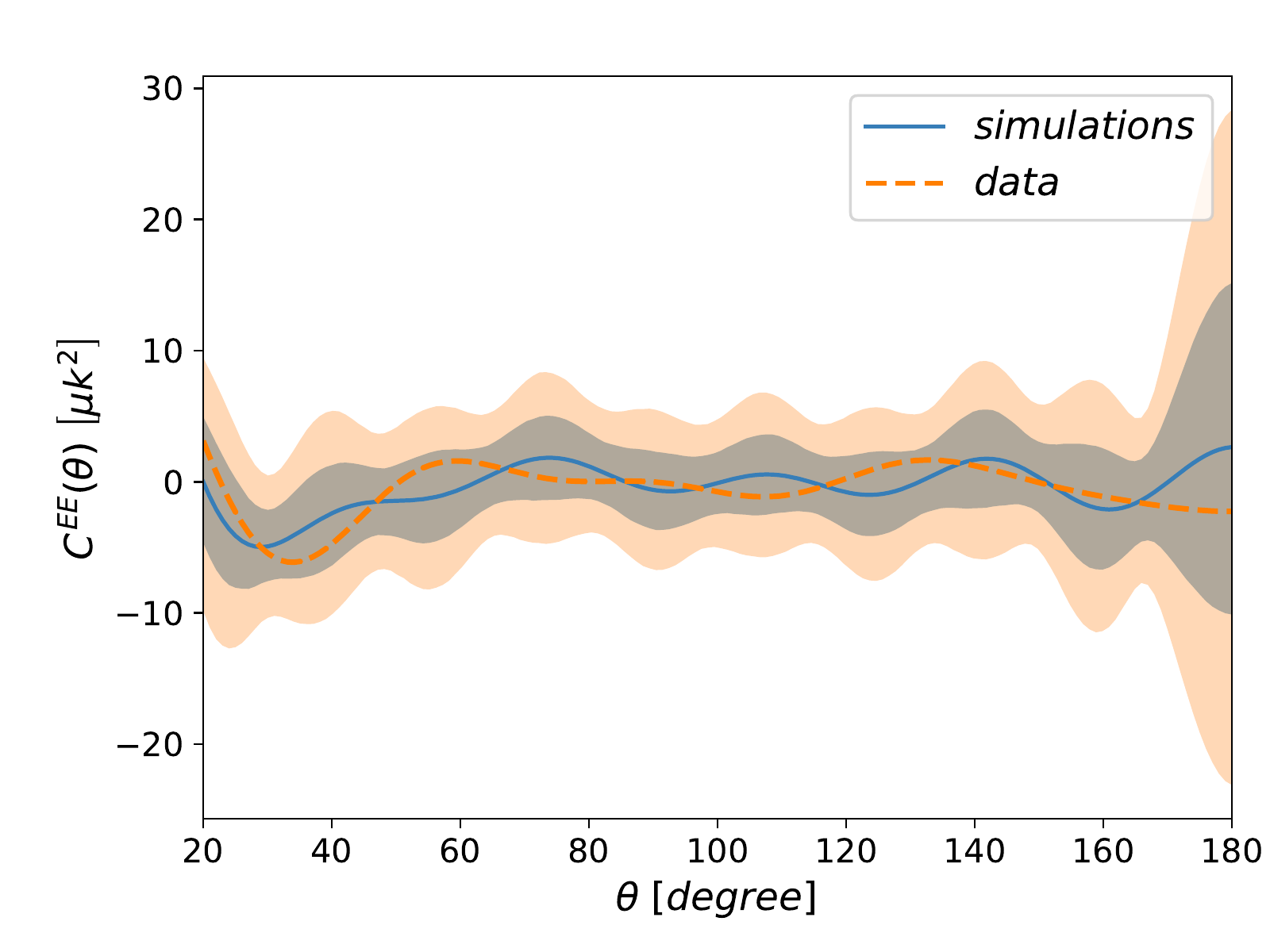}
\hfill
\includegraphics[width=.45\textwidth]{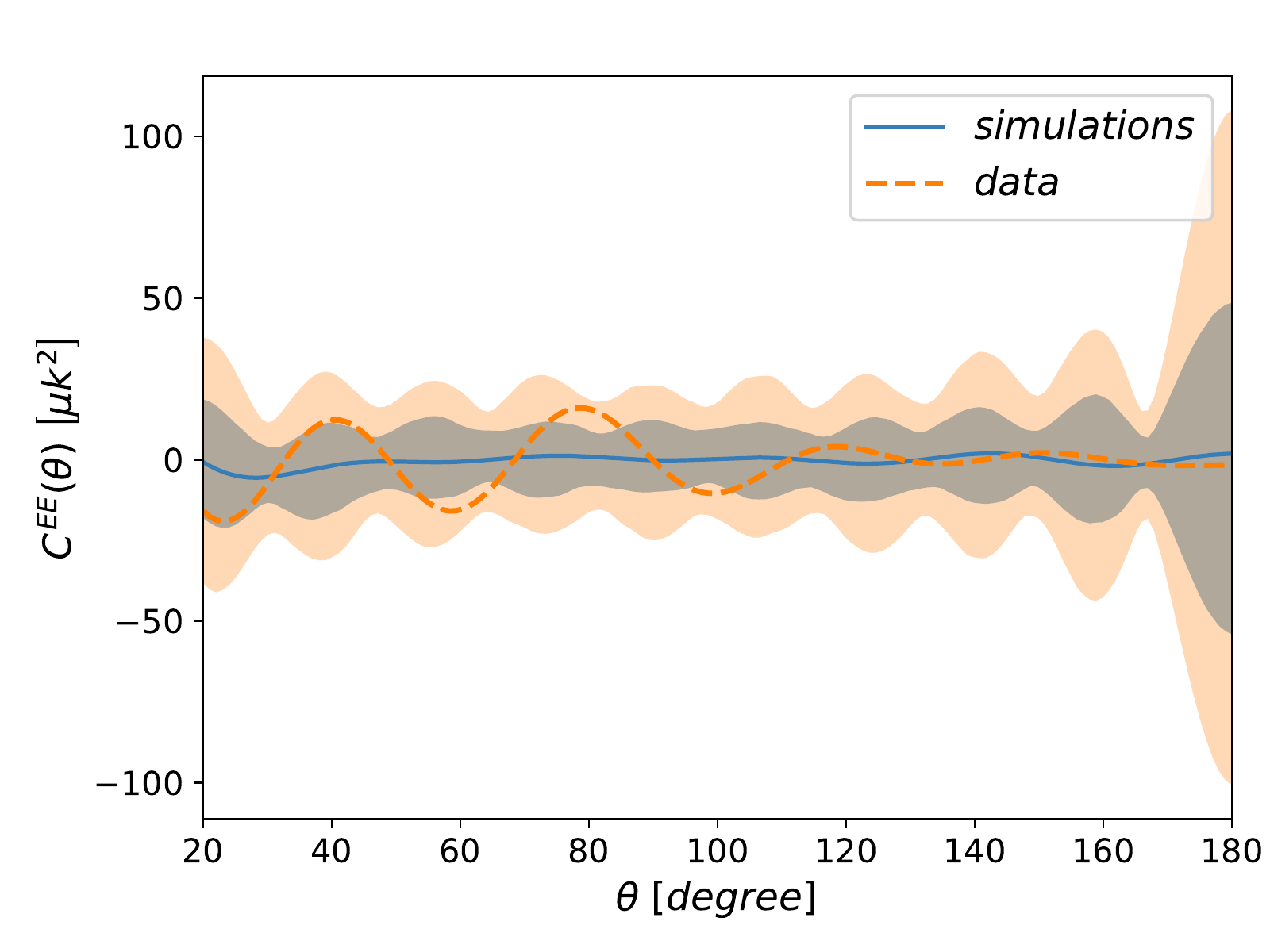}
\caption{\label{fig:cthetaEE} As in Fig.~\ref{fig:cthetaQQ} above but for the two-point angular correlation function built from E-modes. }
\end{figure}

\begin{figure}[tbp]
\centering 
\includegraphics[width=.45\textwidth]{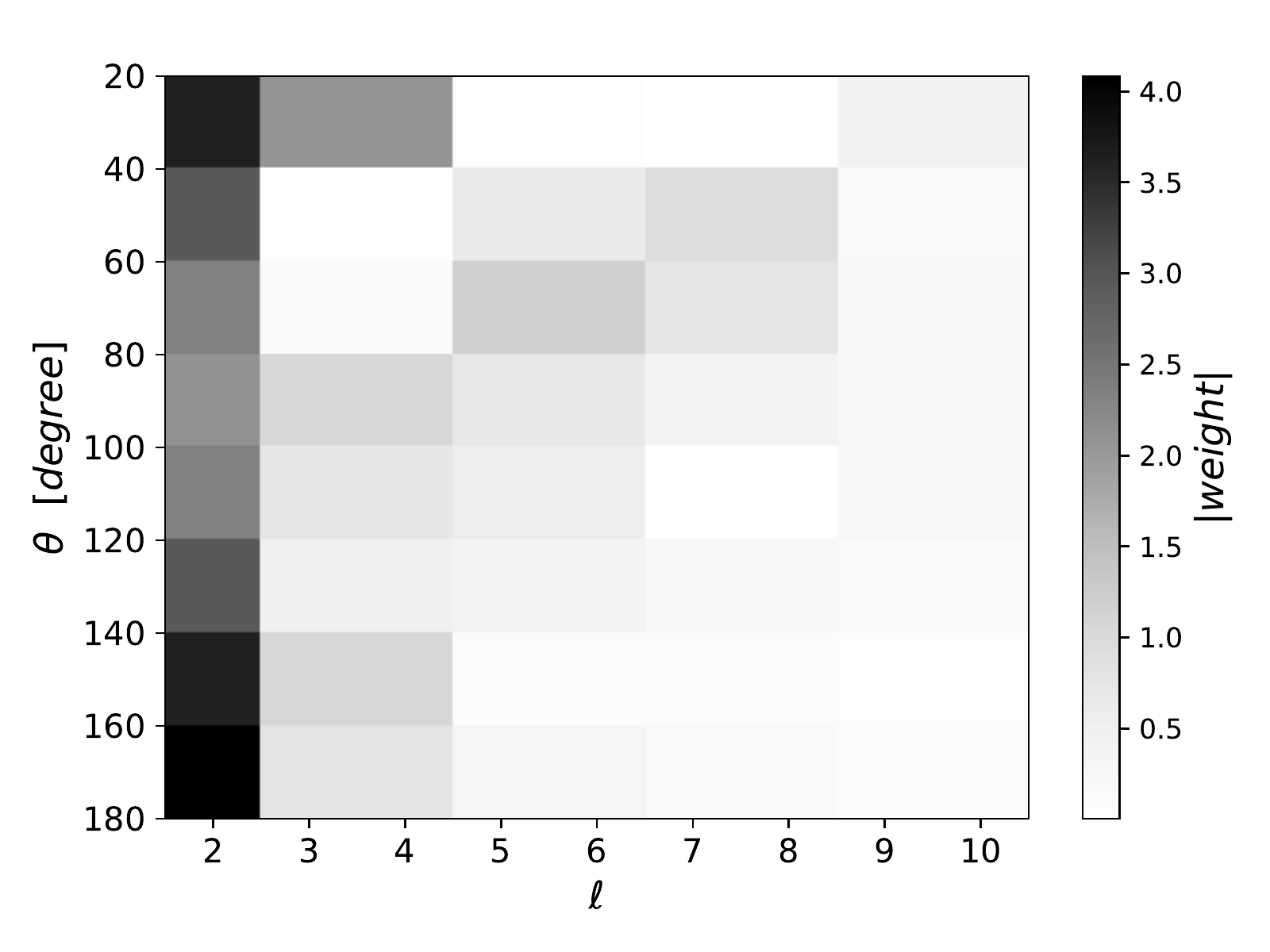}
\hfill
\includegraphics[width=.45\textwidth]{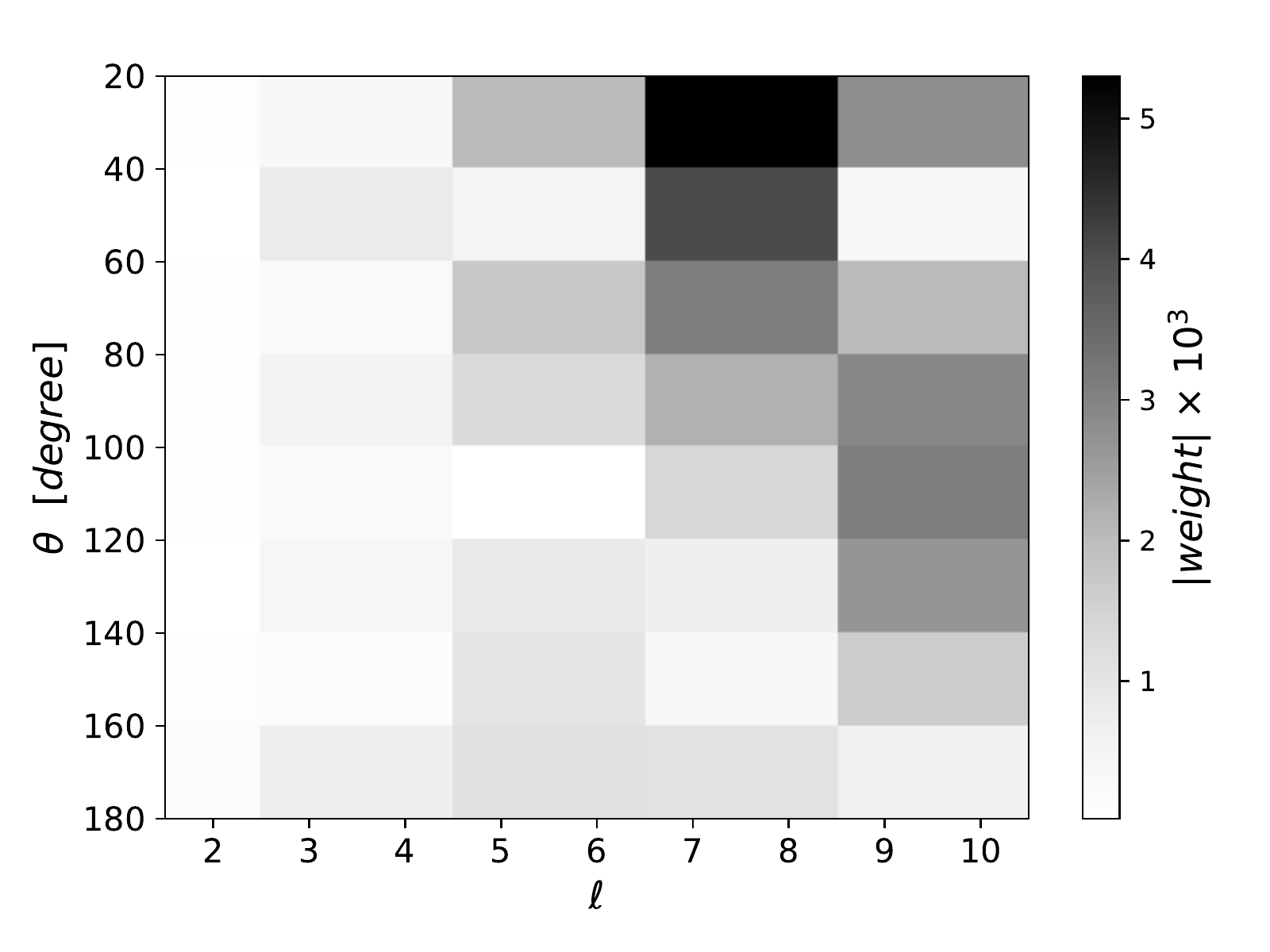}
\caption{\label{fig:pesi} Absolute value of binned geometrical weights applyed to $C^{EE}_{\ell}$s entering in the definition of the Q Stokes parameter (left panel) and of the local E-modes (on the right) correlation functions.}
\end{figure}

\begin{figure}[tbp]
\centering 
\includegraphics[width=.45\textwidth]{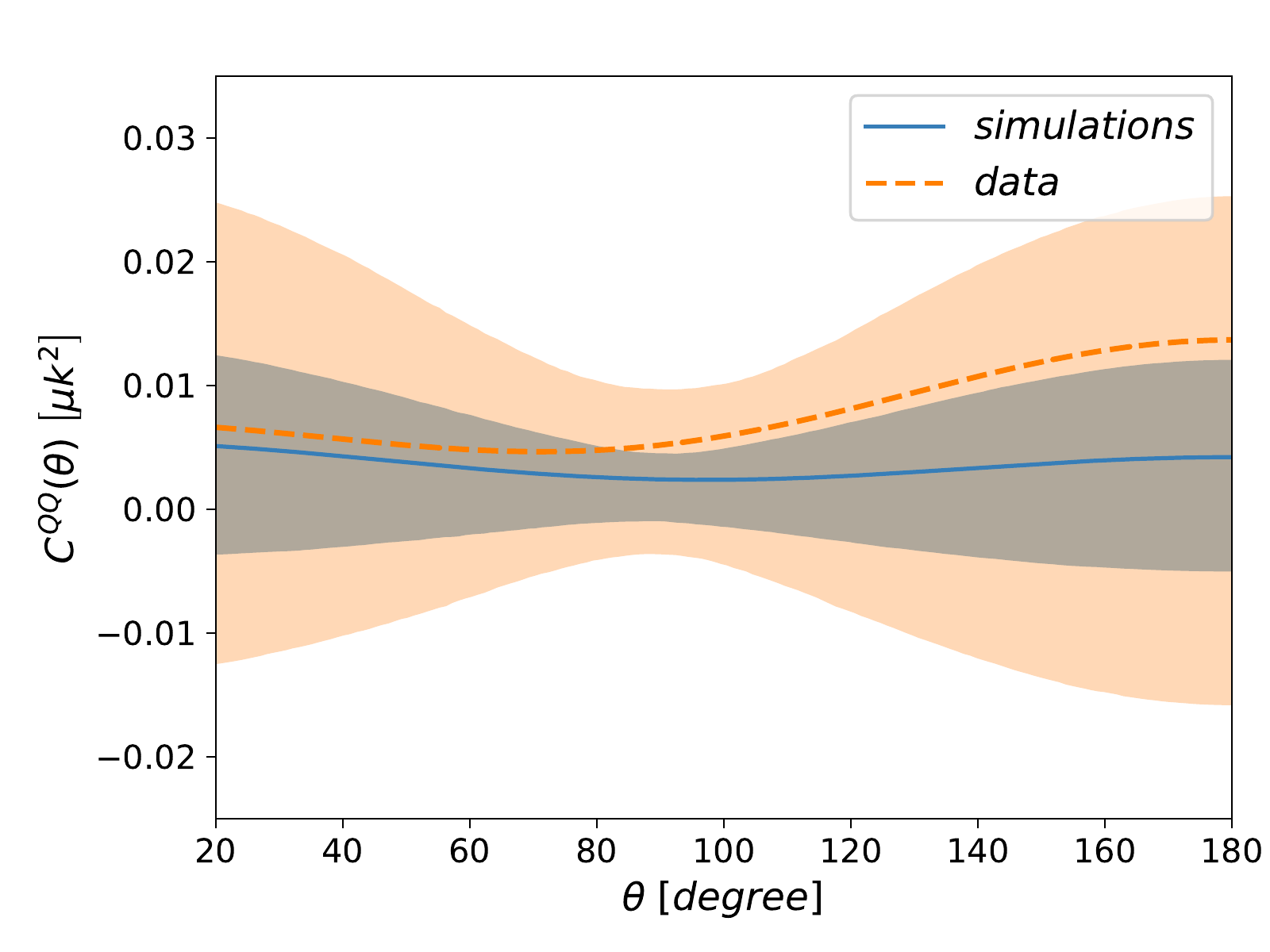}
\hfill
\includegraphics[width=.45\textwidth]{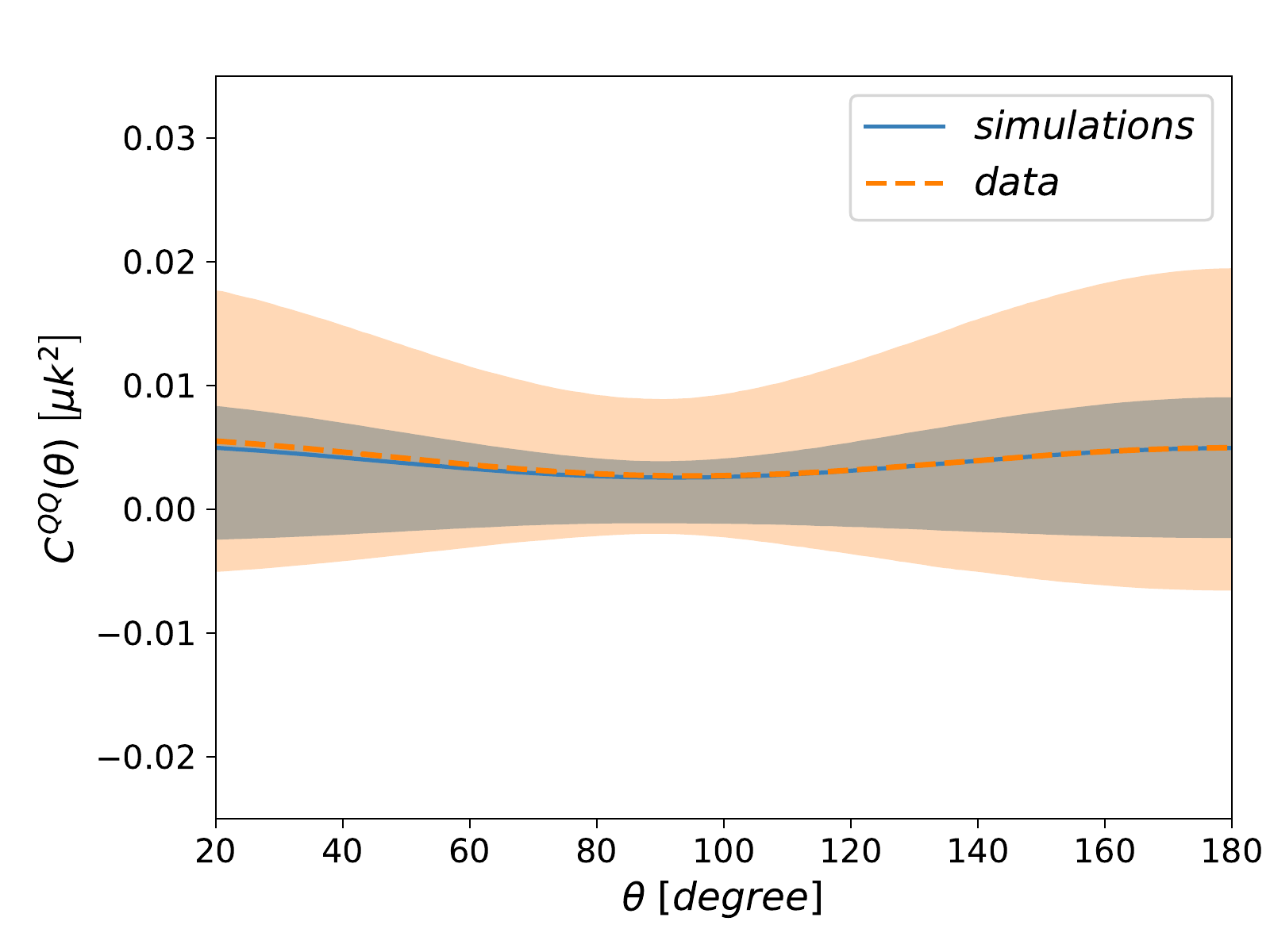}
\caption{\label{fig:cthetaQQquad} Two point angular $Q$ correlation function computed only with the quadrupole contribution. The shaded region represents the $68\%$ and $95\%$ C.L.  The datasets employed are \crossdata\ on the left and \combdata\ on the right.}
\end{figure}

\begin{figure}[tbp]
\centering 
\includegraphics[width=.45\textwidth]{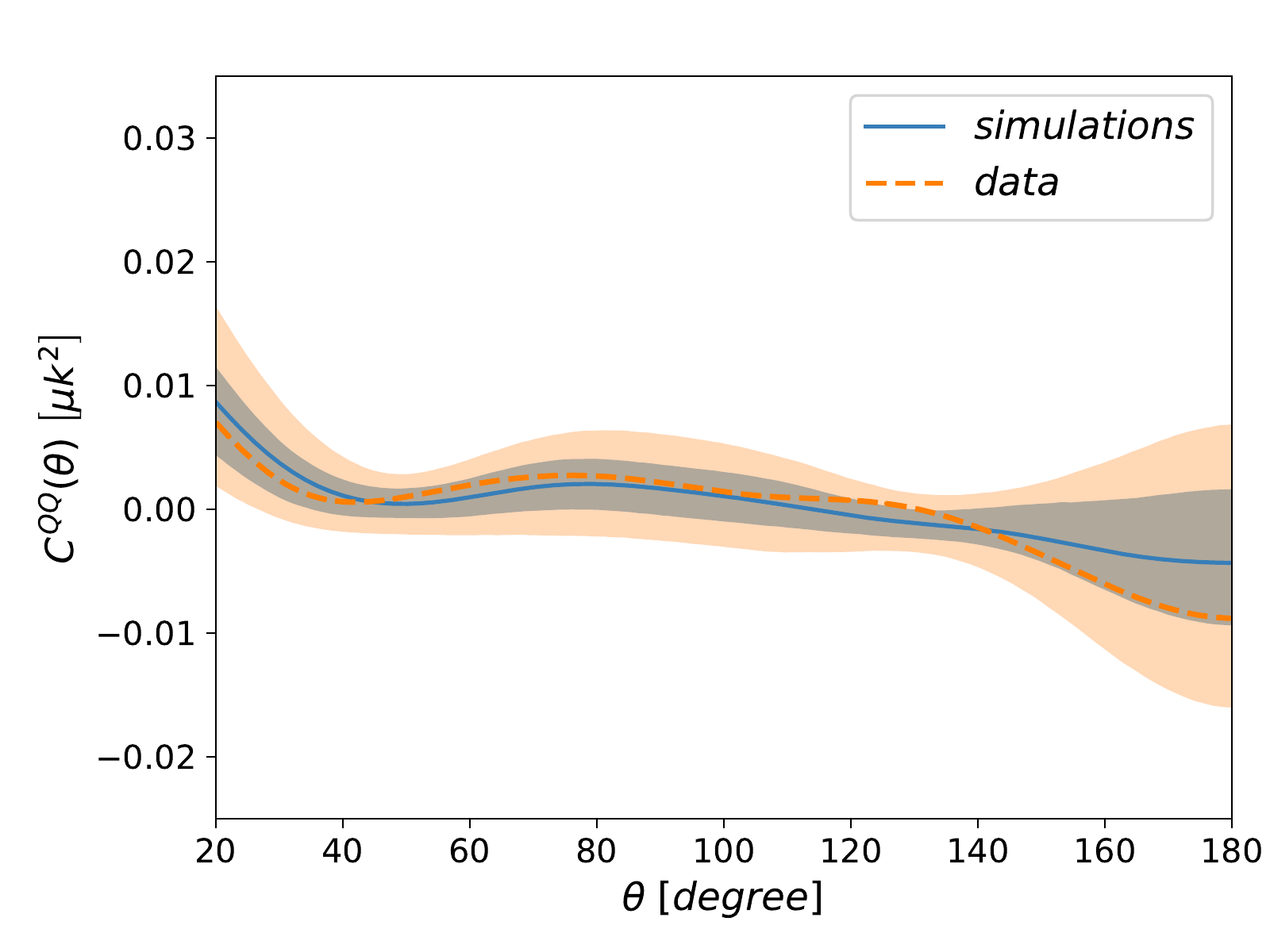}
\hfill
\includegraphics[width=.45\textwidth]{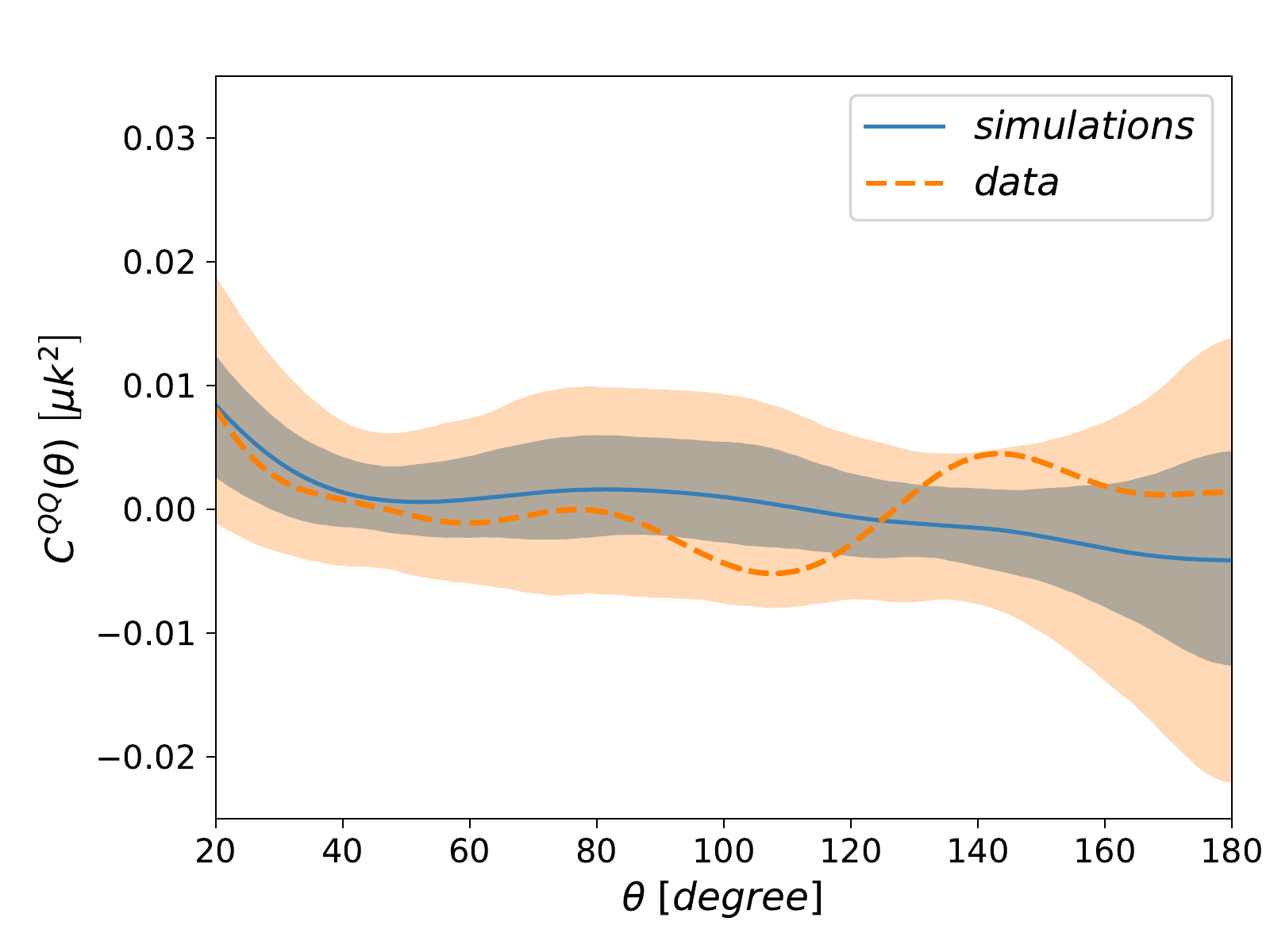}
\caption{\label{fig:cthetaQQnoquad} Two point angular $Q$ correlation function computed starting from the octupole. The shaded region represents the $68\%$ and $95\%$ C.L.The datasets employed are \crossdata\ on the left and \combdata\ on the right. The analysis is performed up to $\ell_{\rm max}=10$}
\end{figure}

We show in Fig.~\ref{fig:cthetaQQ} and \ref{fig:cthetaUU} the $QQ$ and $UU$ angular correlation functions for both the \combdata\ and \crossdata\ datasets, along with mean values and confidence intervals derived from constrained simulations, setting $\ell_{max}=10$.
In Fig.\ \ref{fig:cthetaEE} we show instead the correlation function for local E-modes. Note that only in this latter case the different noise levels of the \combdata\ and \crossdata\ datasets clearly show up in the plots.  Such behaviour can be ascribed to the weights  applied to each multipole when computing the correlation functions out of power spectra. In order to further clarify this aspect we show in Fig.\ \ref{fig:pesi} the geometrical weights of $C^{EE}_{\ell}$ in the definition of Q and local E-modes correlation functions, see eq. \eqref{sub_Gp}:
\begin{subequations}
    \label{eq:weights}
    \begin{align}
    &W_{Q}(\theta) =  \sum_{\ell} \frac{2\ell+1}{4\pi}\left(\frac{2(\ell-2)!}{(\ell+2)!}\right)G^{+}_{\ell 2}(\cos(\theta)),\\
    &W_{\hat{E}}(\theta) = \sum_{\ell} \frac{2\ell+1}{4\pi}\frac{(\ell+2)!}{(\ell-2)!}\mathcal{P}_{\ell}(\cos{\theta})    
    \end{align}
\end{subequations}
Here $G^{+}_{\ell 2}$ is defined as in eq. \eqref{sub_Gp}.
Both quantities entering the definition of the weights (i.e. the angle $\theta$ and the multipole $\ell$) are binned and the plots show the total weight inside each bin. To highlight the contribution of the quadrupole it is shown without applying any binning. 
Fig.\ \ref{fig:pesi} shows how the $Q$ and $U$ correlation functions are dominated by very low multipoles, in particular by the quadrupole, while the correlation function of local E-modes  is more susceptible to variations at high multipoles considered here. This has a clear impact on the variance of the correlation function itself. If the $Q$ correlation function is computed only from the quadrupole, \combdata\ is more sensitive than \crossdata\ (see Fig.\ \ref{fig:cthetaQQquad}), being the latter dominated by residual dipole leakage as shown in Fig.\ \ref{fig:rapp_var}. For all the higher multipoles, instead, \crossdata\ is clearly more constraining (see Fig.\ \ref{fig:cthetaQQnoquad}), partially, but not completely, compensating the quadrupole behaviour. Analogous results can be obtained for $U$ Stokes field. The same multipole split for local $E$ modes does not show the same trend, as the variance of the correlation function remains substantially unchanged if the quadrupole is excluded. 

\begin{figure}[tbp]
\centering 
\includegraphics[width=.45\textwidth]{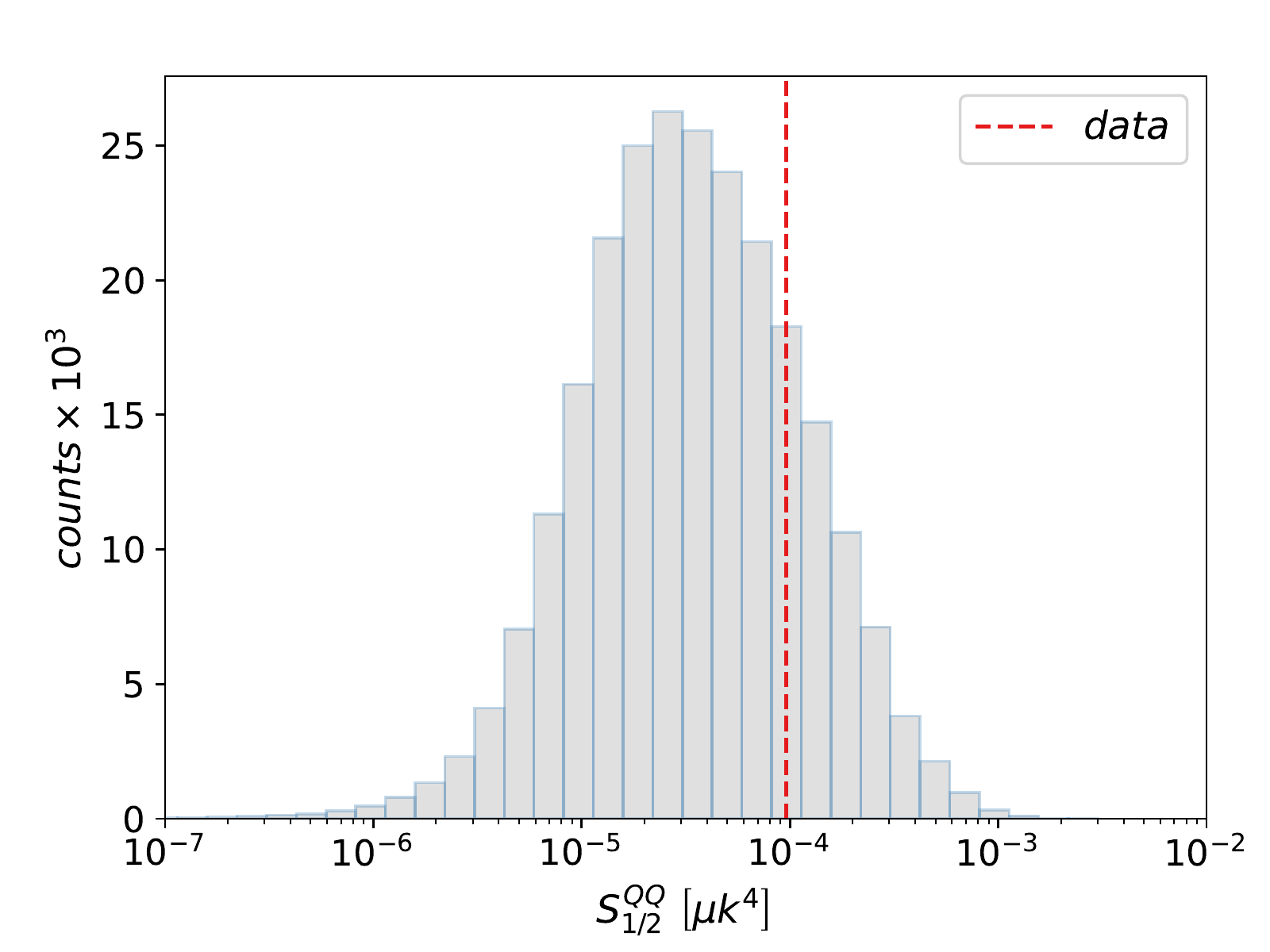}
\hfill
\includegraphics[width=.45\textwidth]{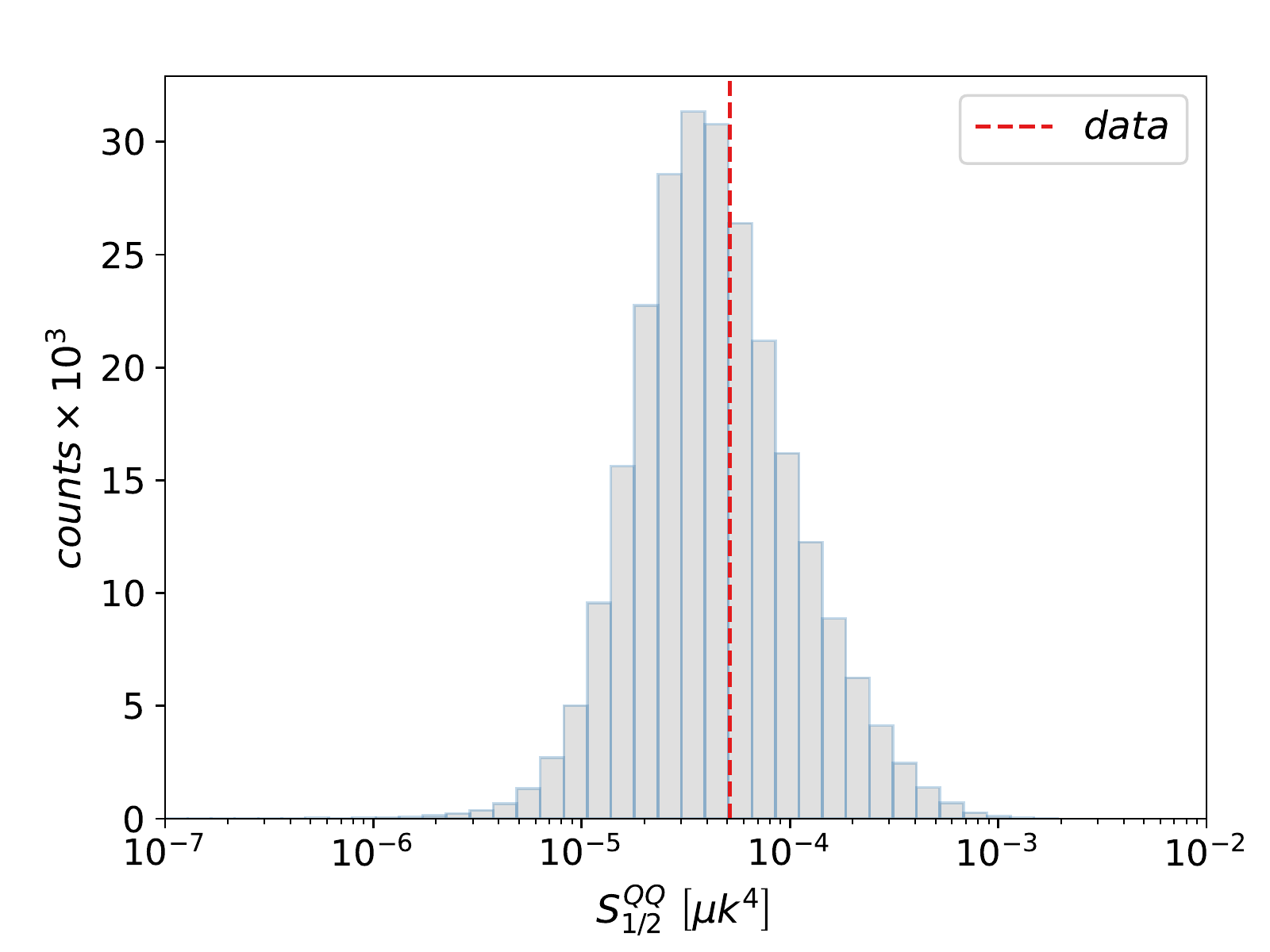}
\caption{\label{fig:statisticaQQ}Empirical distribution for $S_{1/2}$ values from simulations (grey) and data (red). The estimator is computed on $Q$ correlation function. The maximum multipole used is $\ell_{max}= 10$. The left panel is for \crossdata\ dataset and the right one for \combdata\ .}
\end{figure}

\begin{figure}[tbp]
\centering 
\includegraphics[width=.45\textwidth]{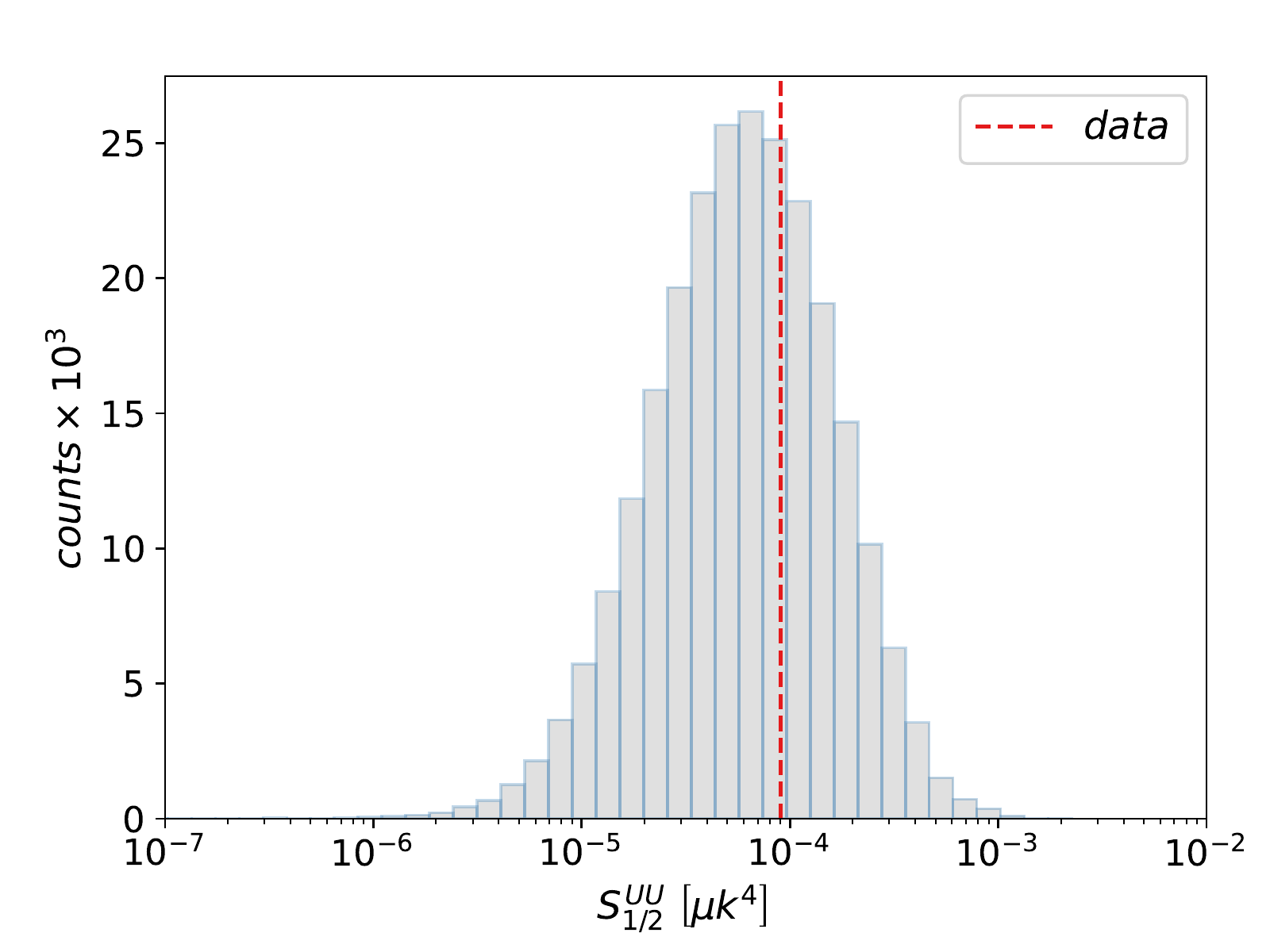}
\hfill
\includegraphics[width=.45\textwidth]{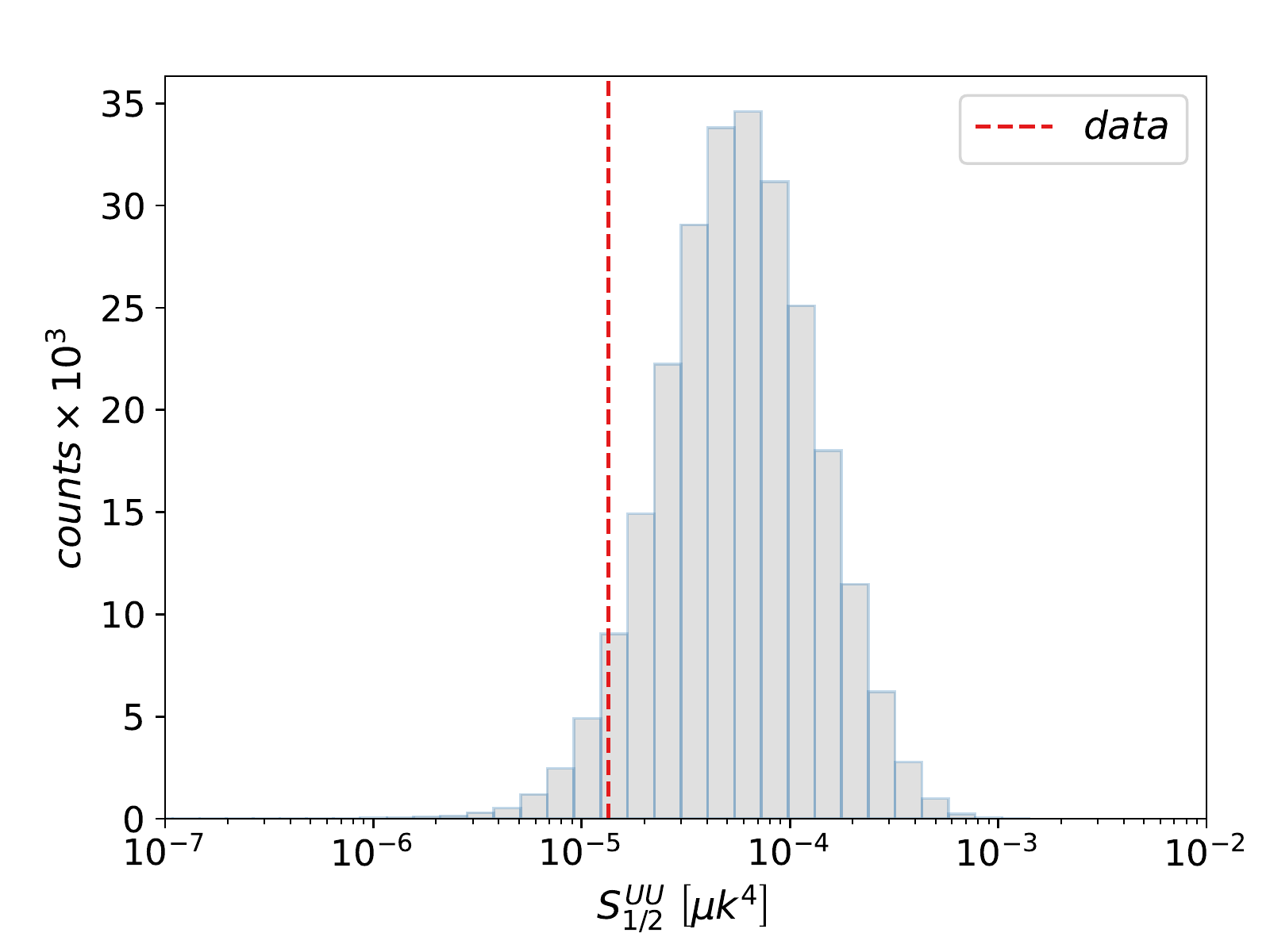}
\caption{\label{fig:statisticaUU} Same as in Figure \ref{fig:statisticaQQ} but for $U$. 
}
\end{figure}

\begin{figure}[tbp]
\centering 
\includegraphics[width=.45\textwidth]{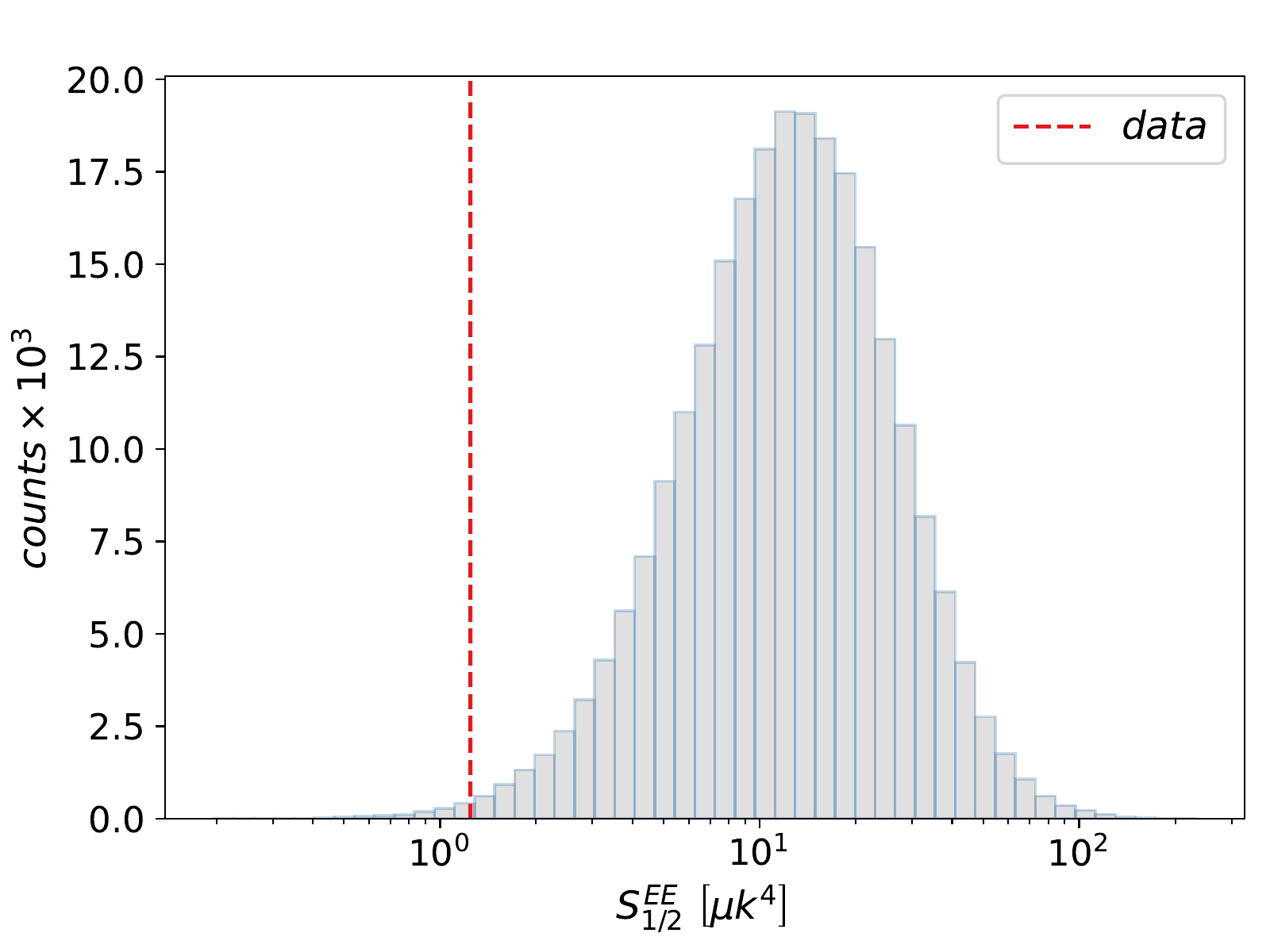}
\hfill
\includegraphics[width=.45\textwidth]{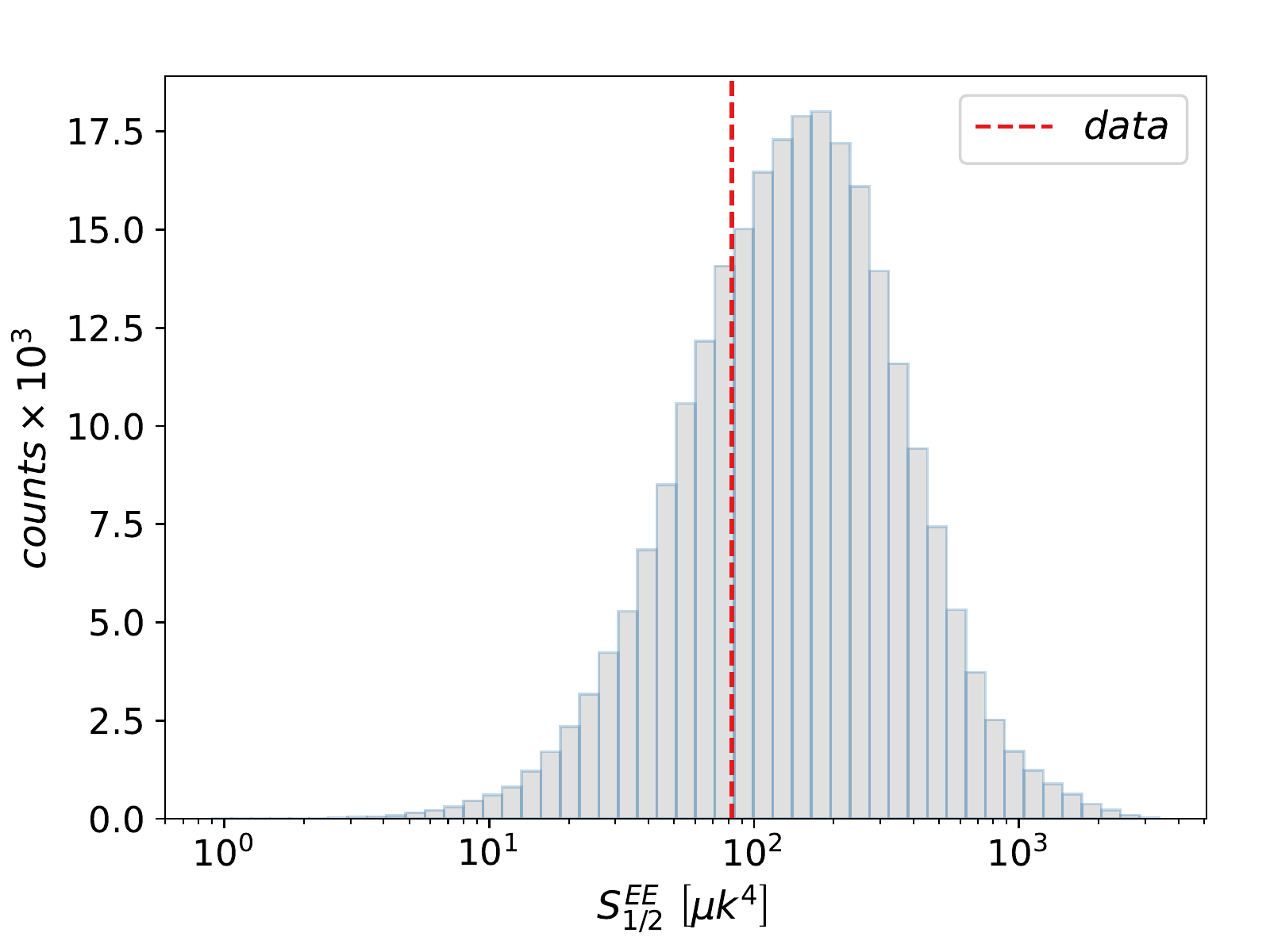}
\caption{\label{fig:statisticaEElog} Same as in Figure \ref{fig:statisticaQQ} but for $E$-modes.} 
\end{figure}

In figure \ref{fig:statisticaQQ}, \ref{fig:statisticaUU} and \ref{fig:statisticaEElog} we plot in light grey the distribution of the $S_{1/2}$ estimators for the Q and U fields respectively and for the local E modes, as defined in eq.~(\ref{eq:shalfQQ}) and (\ref{eq:shalfXX}). The red line represents the value of the estimators on data. In all figures the panel on the left refers to the \crossdata\ dataset and the one on the right to the \combdata\ dataset.
Specifically, we compute the integrals in eq.~(\ref{matrixTT}) and (\ref{eq:spin2matrix}), involved in the computation of $I_{\ell\ell'}$ matrices, in the angular range $60^{\circ}-180^{\circ}$, coherently with previous analysis (see \cite{Spergel:2003cb}, \cite{Akrami:2019bkn}). 

\begin{table}[!h]
\begin{center}
\begin{tabular}{l|c|c|c|c}
     & $S_{1/2}^{data}$ & $S_{1/2} >$ $S_{1/2}^{data}$ & $S_{1/2}^{data_{\ell>2}}$ & $S_{1/2}^{\ell>2} >$ $S_{1/2}^{data_{\ell>2}}$  [$\%$]\\
    \hline
    & [$\mu$K$^{4}$] & [$\%$] & [$\mu$K$^{4}$] & [$\%$]\\
    \hline
     EE \crossdata\ & 1.25 & 99.5 (99.6) &1.30 & 99.5 (99.6)\\
    \hline 
     EE \combdata\ & 82.4 & 71.6 (72.3)& 82.7 & 71.5 (72.2)\\
    \hline
    \hline
    & $10^{5}$ $\times$ [$\mu$K$^{4}$] & [$\%$] & $10^{5}$ $\times$ [$\mu$K$^{4}$] & [$\%$]\\
    \hline
     QQ \crossdata\ & 9.57 & 19.6 (31.7)& 0.85 & 48.8 (49.0)\\
    \hline
     UU \crossdata\ & 9.0 & 34.2 (41.8)& 3.21 & 49.2 (53.3)\\ 
    \hline
     QQ \combdata\  & 5.14 & 39.4 (50.2)& 3.33 & 25.5 (27.3)\\
    \hline
     UU \combdata\  & 1.35 & 95.0 (95.6)& 2.25  & 75.2 (77.6)\\
    \hline
\end{tabular}
\end{center}
\caption{$S_{1/2}$ measured on data (second column including the quadrupole term and fourth column excluding the quadrupole term) and percentage of simulations with value of the estimator larger than the one found on data (third column including the quadrupole term and fifth column excluding the quadrupole term). The values reported in brackets refer to unconstrained simulations. The sensitivity associated to the percentage of simulations with $S_{1/2}$ higher than data is $\sim0.2\%$.}
\label{tab:results}
\end{table}

In Table \ref{tab:results} we report the value of $S_{1/2}$ estimator on data for the analysed datasets, both with and without the quadrupole contribution. We also show the percentage of simulations having a value of $S_{1/2}$ larger than data, the values reported in brackets refer to the unconstrained simulations case.
As previously discussed, the local E modes estimator does not sizeably depend on the inclusion of the quadrupole, differently from what happens for $Q$ and $U$. As the \crossdata\ dataset has a higher signal to noise ratio with respect to \combdata, we rely on it to make an assessment on the significance of the anomaly for E-modes. 
The value of the estimator on data is $S_{1/2}^{data} = 1.25~\mu k$ for \crossdata, with a lower tail probability of $0.5\%$, which suggests a low power in polarization data up to $\ell =10$. 
As previously noted, the correlation functions for $Q$ and $U$ change significantly when the quadrupole is excluded from the analysis. 
The higher variance due to residual dipole on the quadrupole in the \crossdata\ dataset suggests to base our considerations on the \combdata\ dataset. The trend of the latter dataset seems to indicate that the contribution of $\ell=2$ increase the power of the low multipoles in the data with respect to simulations. This effect is highlighted  by the decrease of the percentage of simulations with a value of $S_{1/2}^{data}$ higher than data when excluding the quadrupole from the analysis (compare III and V columns in table \ref{tab:results} for the QQ case of \combdata\ dataset). This conclusion holds for both constrained and unconstrained simulations.

\begin{figure}[tbp]
\centering 
\includegraphics[scale = 0.50]{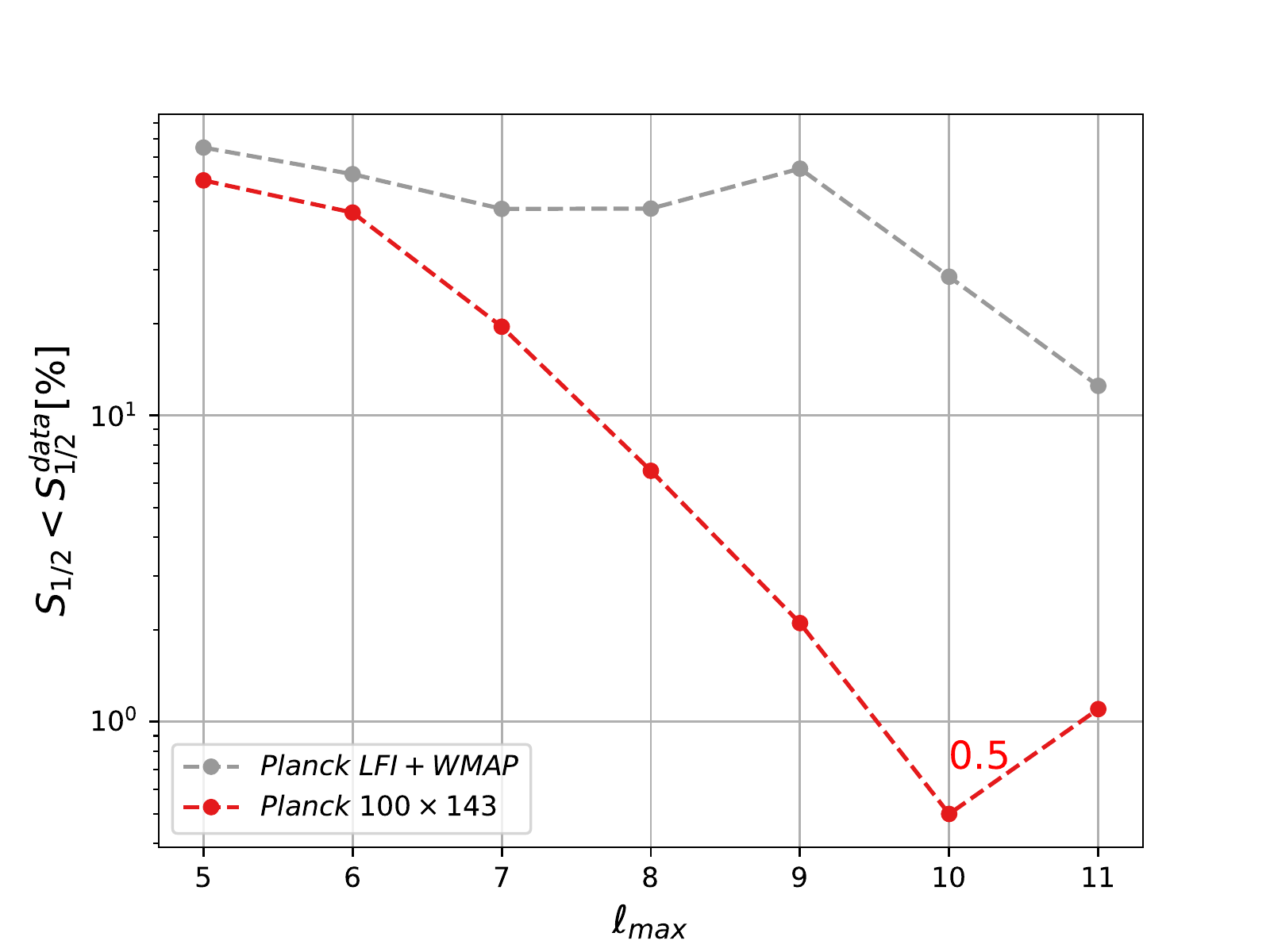}
\caption{\label{fig:lowertail} Lower tail probability for the local E modes estimator computed on data w.r.t constrained simulations for both \crossdata\ (red curve) and \combdata\ (grey curve) datasets. The region of multipoles higher than $11$ is excluded from the plot as considered dominated by noise. }
\end{figure}

To further investigate the contribution of each multipole in determining the relative power of data with respect to the empirical distributions, we computed the value of the estimators on constrained simulations and data gradually increasing the maximum multipole included in the analysis. As expected, the quadrupole dominates the $Q$ and $U$ results which therefore cannot be taken as representative of the behaviour of the other low multipoles. 
On the contrary, the local E-modes are not very sensitive to the quadrupole and exhibit a decreasing lower tail probability as a function of $\ell_{max}$ as shown in Figure \ref{fig:lowertail}.
This is true for the \crossdata\ dataset but also for the \combdata\ dataset pending some scattering which can be ascribed to the lower signal-to-noise of the latter. 
For multipoles above $\ell \sim 10$ both datasets are noise-dominated and any trend is lost. Taking here the \crossdata\ dataset as our benchmark, due to its higher signal-to-noise, we find a lower-tail probability of $0.5\%$. Note that, in spite of the fact that, the quadrupole for this dataset exhibits extra variance, as explained above, yet we choose it as our benchmark given the fairly low sensitivity of the local E-modes $S_{1/2}$ estimator to the value of the quadrupole itself.

\begin{figure}[tbp]
\centering 
\includegraphics[scale = 0.50]{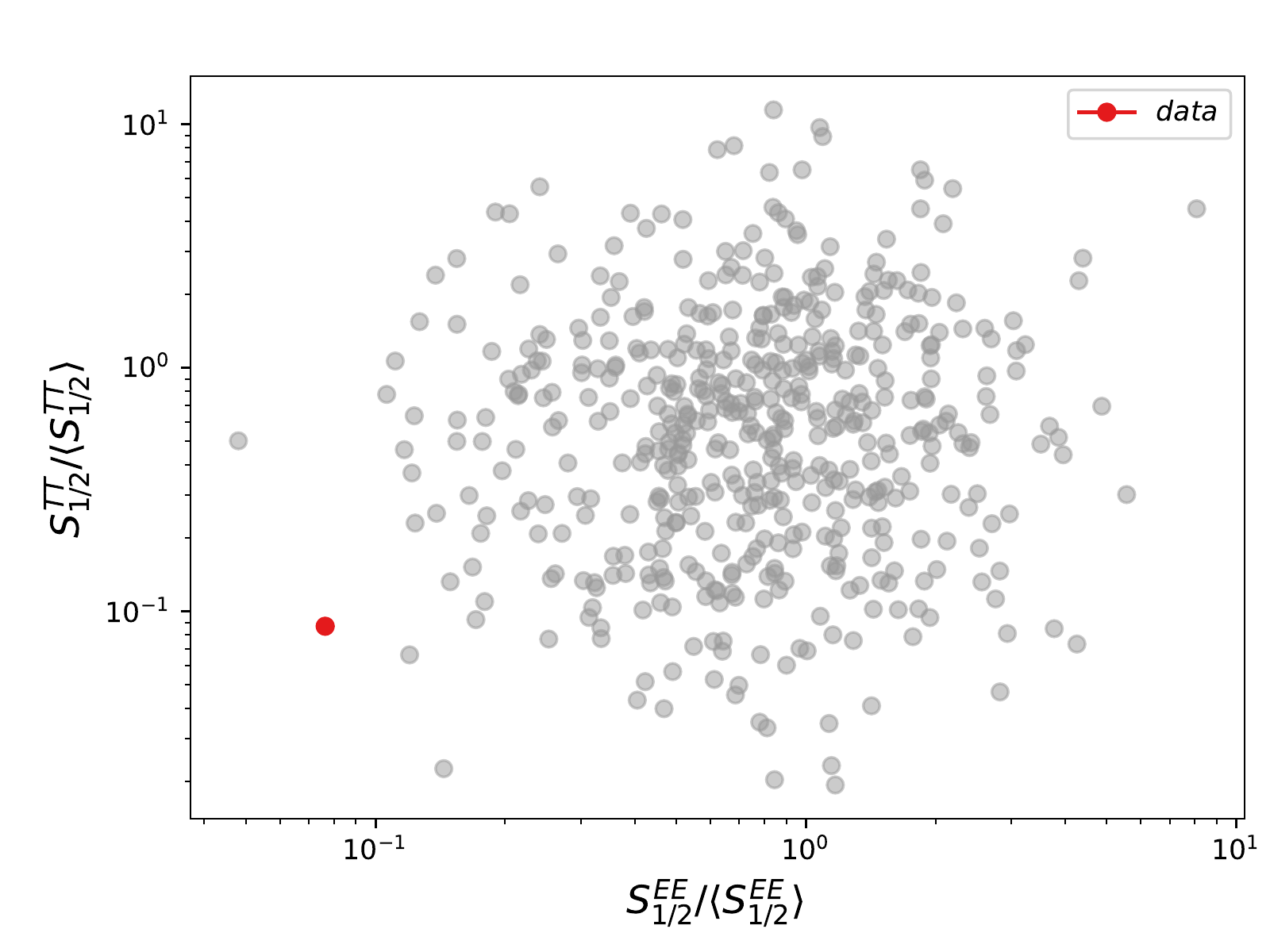}
\caption{\label{fig:scatter} Joint behaviour of the $S_{1/2}$ estimator in temperature and polarization for $500$ unconstrained simulations (grey dots) and data (red dot) of \crossdata\ dataset. All values of the estimator in temperature and polarization reported here are normalized to the empirical mean of the corresponding distribution. The position of data with respect to the simulations suggest a low combined power of T and E correlation functions.}
\end{figure}

As a final test, we study the joint distribution of the $S_{1/2}$ estimators in temperature and polarization. In Figure \ref{fig:scatter}, we show the distribution in the ($S_{1/2}^{EE},\,S_{1/2}^{TT}$) plane of $N_\mathrm{sims} = 500$ unconstrained  \crossdata\ simulations (grey dots), compared to values computed on the actual \crossdata\ data (red dot). For each simulation the value of the estimator in temperature is normalized to the empirical mean of the $S_{1/2}^{TT}$ simulations distributions and the value of the estimator in polarization is normalized to the empirical mean of the $S_{1/2}^{EE}$ simulations distributions. The same normalization is applied to data. 

\begin{figure}[tbp]
\centering 
\includegraphics[scale = 0.50]{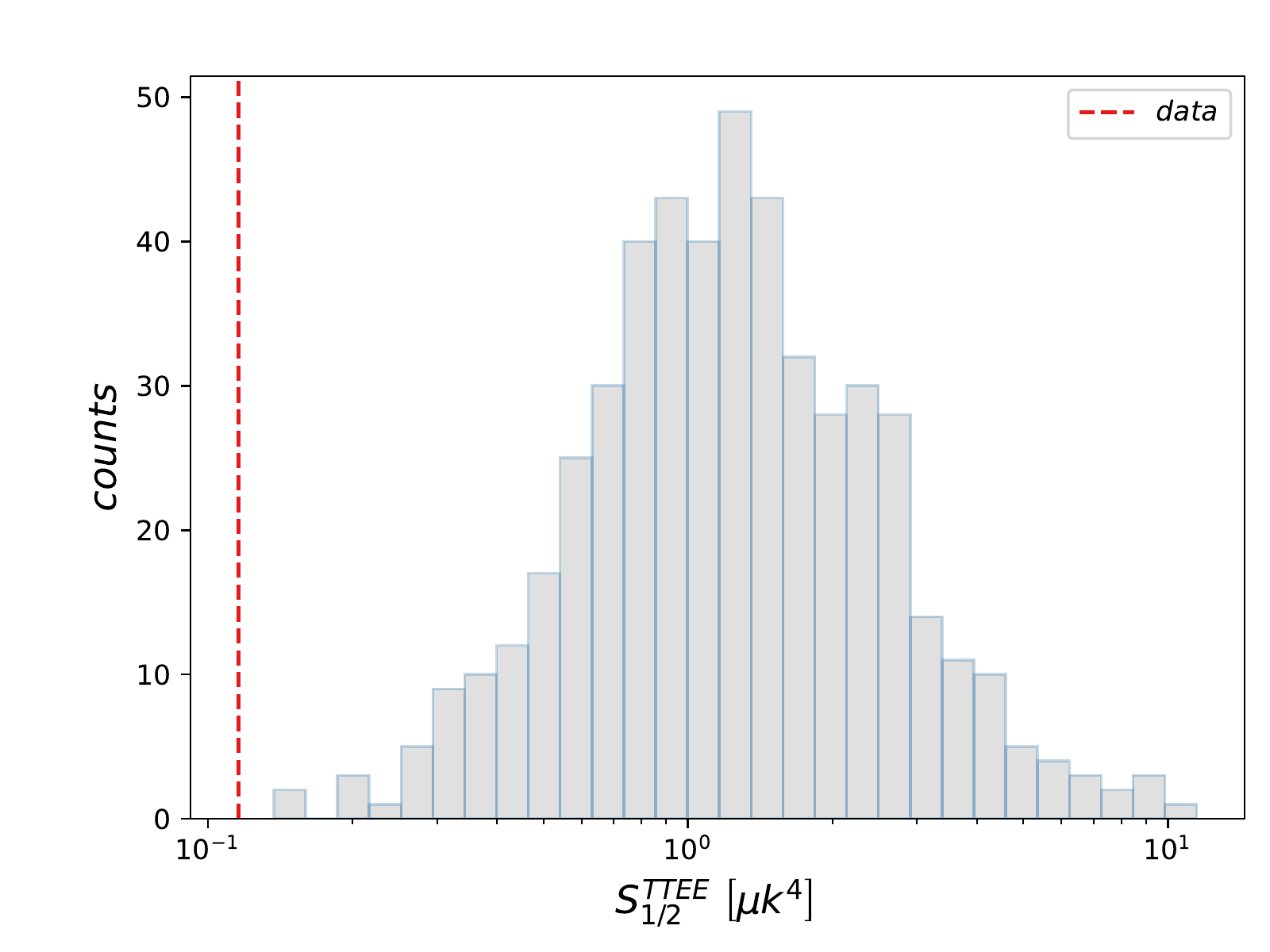}
\caption{\label{fig:distance} Distribution of the $S_{1/2}^{EE,TT}$ estimator for $500$ \crossdata\ simulations (grey) and data (red vertical line). The upper limit of the lower tail probability of the estimator computed on simulations with respect to data is $0.2\%.$}
\end{figure}

Given the relatively small number of simulations available, we seek to compress the information encoded in the two-dimensional joint distribution into a single, one-dimensional estimator. To this purpose, we extend the use of the $S_{1/2}$ estimator from one to two dimensions, computing the distance of the points from zero as in formula \ref{distance}:

\begin{equation}
    S_{1/2}^{EE,TT} = \sqrt{\left(\frac{S_{1/2}^{TT}}{\langle S_{1/2}^{TT}\rangle}\right)^{2}+\left(\frac{S_{1/2}^{EE}}{\langle S_{1/2}^{EE}\rangle}\right)^{2}}
    \label{distance}
\end{equation}
The resulting distribution is shown in figure \ref{fig:distance}.
We find that no simulations have values of $S_{1/2}^{EE,TT}$ lower than data. Given that the sensitivity associated to our Monte Carlo is $1/N_\mathrm{sims} = 0.002$, we can quote this number as an upper limit to the lower tail probability associated to the $S_{1/2}^{EE,TT}$ value measured on the data.

\section{Conclusions}
\label{sec:conclusions}

In this paper we extend the study of the lack-of-power anomaly to the CMB polarization field, analysing the most constraining large-scale datasets currently available, which are the \combdata\ dataset \cite{Natale:2020owc} and the \crossdata\ dataset \cite{Delouis:2019bub}.
Adopting a frequentist approach, we assume \planck\ 2018 + \srolltwo\ cosmological fiducial model \cite{Pagano:2019tci} and in particular a specific value for reionization optical depth $\tau = 0.0591$ which is an important choice for the angular scales probed. The value of $\tau$ used is also compatible with that obtained by \cite{Natale:2020owc} when using the WMAP$+$LFI dataset in polarization, together with the \commander\ 2018 solution in temperature.
We employ the $S_{1/2}$ estimator \cite{Yoho:2015bla}, which is based on the two-point correlation function of $Q$ and $U$, see eq.~(\ref{eq:shalfQQ}), and of the local $E$-modes, see eq.~(\ref{eq:shalfXX}). We compute such estimators on both data and realistic simulations, which contain signal, noise and residual systematic effects, and compare empirical distributions from simulations with data results. 
We employ fully polarised signal, considering both Planck CMB temperature constrained and unconstrained simulations, and limit most of the analysis to $\ell_{max}=10$, a multipole above which both datasets are fully noise-dominated.
We calculate the correlation function for $Q$ and $U$ and show that it is largely dominated by the quadrupole. This clearly impacts the results obtained, which show negligible variation when the maximum multipole included in the analysis is varied. 

For both datasets considered, we do not see any anomalous behaviour, except for a mild $2\sigma$ anomaly in the case of \combdata\ $U$ correlation function. This suggests that the power in the very low multipoles, in particular in the quadrupole, is not anomalously low in data. 
It is worth noting however, that both datasets include non negligible uncertainties at $\ell=2$, mostly of systematic origin for \crossdata\ and mostly statistical for \combdata, these uncertainties likely affect the constraining power of a possible polarisation anomaly.

On the other hand the estimator involving local E-modes behaves differently being more sensitive to the $\ell_{max}$ used in the analysis, and thus giving information on the integrated power of the lower multipoles. In Figure \ref{fig:lowertail} we see that the lower tail probability for the two datasets follows the same descending trend and in particular the \crossdata\ seems to suggest a low power in data with respect to the simulations considered.  

The behaviour of the $S_{1/2}$ estimators on Q, U and local E-modes in the case of unconstrained simulations appears to be similar to what described for the constrained case. In particular the lower tail probability for $\ell_{max} = 10$ of the \crossdata\ dataset is $0.4\%$, indicating again a low power of data with respect to simulations.

These results suggest that the large-angle CMB polarisation data behave in a similar way to temperature, exhibiting a mild low power anomaly, presumably originating not only from the quadrupole but rather than from the combined behaviour of all the multipoles $\ell\le10$.

This conclusion is strengthened  by the analysis of the joint behaviour of the $S_{1/2}$ estimators in temperature and polarization. We employ $500$ unconstrained simulations of the \crossdata\ dataset and find that the lower tail probability associated to the value of the joint estimator measured on the data is $<0.2\%$, thus confirming the anomalous behaviour of the data with respect to the expectations in the framework of the $\Lambda$CDM model.

\begin{figure}[tbp]
\centering 
\includegraphics[scale = 0.50]{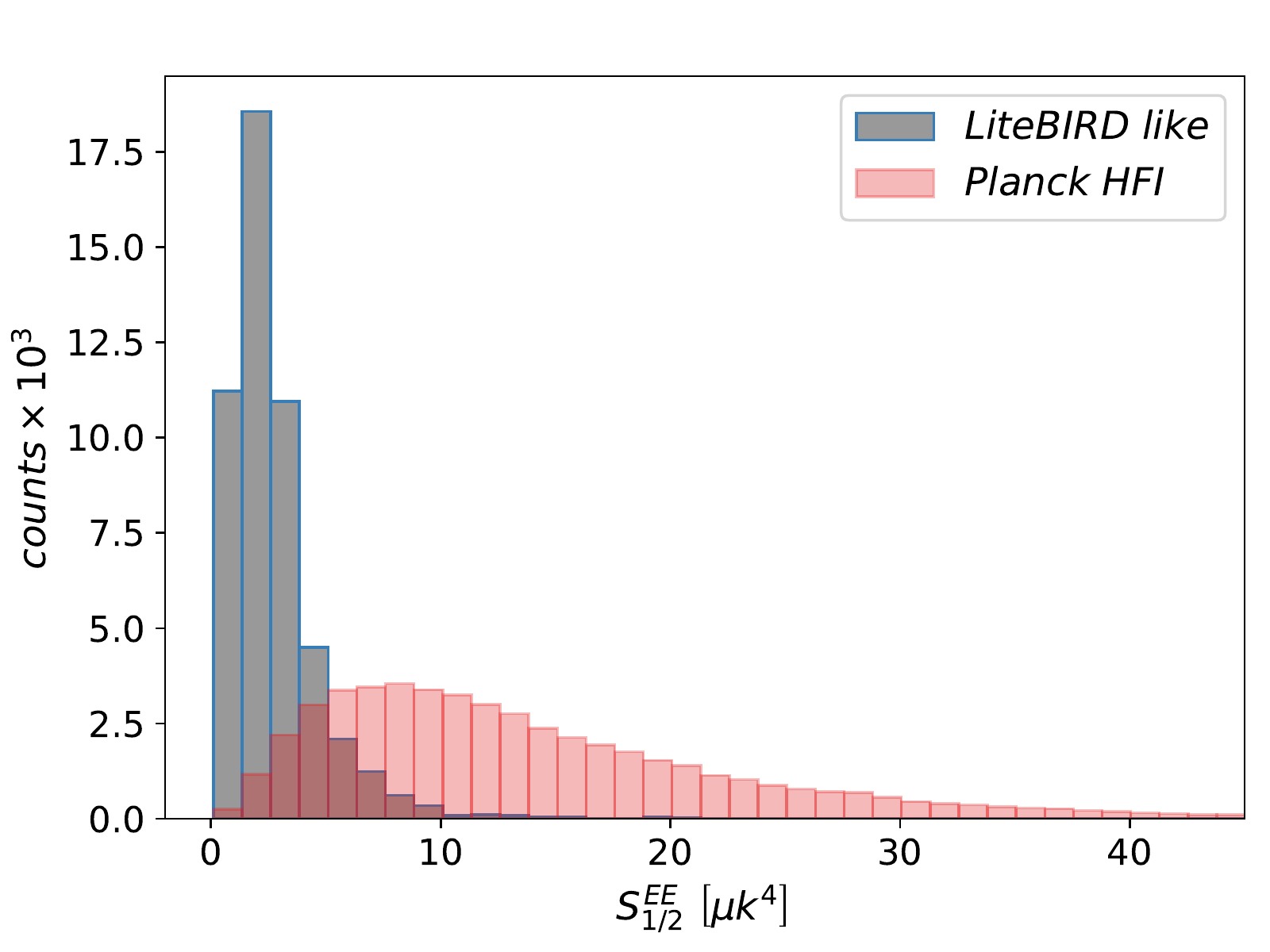}
\caption{\label{fig:LB_planck}Distributions comparison between LiteBIRD like experiment (grey) and \crossdata\ dataset (magenta) simulations. The lower width and the left shift of the peak of the grey distribution are due to the lower level of noise which turns into an increase of constraining power for LiteBIRD like experiment.}
\end{figure}

The present analysis has been carried out with the best datasets currently available at large angular scales, which are however limited by the still significant amount of noise in polarisation observations.
This issue will be hopefully overcome by the advent of new data, such as those from LiteBIRD \cite{Sugai:2020pjw}, which are expected to be cosmic variance limited at all scales. 
Through a rough estimate of the noise level from the three most sensitive LiteBIRD bands we expect an increase of five times the constraining power of this test in the case of polarization E modes, see Fig.\ \ref{fig:LB_planck}.
The perspectives for shedding light on this subject are thus high.

\acknowledgments
This work is based on observations obtained with Planck (http://www.esa.int/Planck), an ESA science mission with instruments and contributions directly funded by ESA Member States, NASA, and Canada.
We acknowledge the use of computing facilities at NERSC and those provided by the INFN theory group (I.S.\ InDark) at CINECA. 
Some of the results in this paper have been derived using the following packages: \texttt{healpy} \cite{Zonca:2019vzt,Gorski:2004by}, \texttt{NumPy} \cite{Harris:2020xlr}, \texttt{SciPy} \cite{Virtanen:2019joe}, and \texttt{Matplotlib} \cite{Hunter:2007ouj}.
We acknowledge the financial support from the INFN InDark project and from the COSMOS network (www.cosmosnet.it) through the ASI (Italian Space Agency) Grants 2016-24-H.0 and 2016-24-H.1-2018.

\appendix

\section{Appendix: calculation of $S^{QQ}_{1/2}$ and $S^{UU}_{1/2}$ in terms of power spectrum}
\label{a1}
Writing $S^{QQ}_{1/2}$ and $S^{UU}_{1/2}$ in terms of power spectrum is useful to ease computation. To obtain an analytic expression of these estimators we express the $G^{\pm}_{\ell}(\cos(\theta))$ functions in terms of the reduced Wigner matrices. 
We define
\begin{subequations}
    \label{Dfunc}
    \begin{align}
    &D^{+}_{\ell} = \frac{4(\ell-2)!}{(\ell+2)!}G^{+}_{\ell 2}(\cos\theta);\\
    &D^{-}_{\ell} = \frac{4(\ell-2)!}{(\ell+2)!}G^{-}_{\ell 2}(\cos\theta),
    \end{align}
\end{subequations}
and rewrite the $S^{QQ}_{1/2}$ statistic in terms of the power spectrum in the following way
\begin{equation}
    \begin{split}
    S^{QQ}_{1/2} &\equiv \int_{-1}^{1/2}d(\cos{\theta})[C^{QQ}(\theta)]^{2}\\ 
    &= \sum_{\ell=2}^{\ell_{max}}  \frac{2\ell+1}{8\pi} \frac{2\ell'+1}{8\pi} \int_{-1}^{1/2}d(\cos{\theta})[D^{+}_{\ell}C_{\ell}^{EE}+D^{-}_{\ell}C_{\ell}^{BB}][D^{+}_{\ell'}C_{\ell'}^{EE}+D^{-}_{\ell'}C_{\ell'}^{BB}].
    \end{split}
\end{equation}
Going further in the calculation we obtain:
\begin{multline}
        S^{QQ}_{1/2} = \sum_{\ell=2}^{\ell_{max}}  \frac{2\ell+1}{8\pi} \frac{2\ell'+1}{8\pi}\biggl(C^{EE}_{\ell}C^{EE}_{\ell'} I^{(1)}_{\ell\ell'} 
    +C^{BB}_{\ell}C^{BB}_{\ell'}I^{(3)}_{\ell\ell'} +C^{EE}_{\ell}C^{BB}_{\ell'}I^{(2)}_{\ell\ell'} 
    +C^{BB}_{\ell}C^{EE}_{\ell'}I^{(4)}_{\ell\ell'} \biggr),
\end{multline}
where the $I^{(X)}_{\ell\ell'}$ matrices are defined as:
\begin{equation}
    \begin{split}
    &I^{(1)}_{\ell\ell'} = \int_{-1}^{1/2}d(\cos{\theta})D^{+}_{\ell}D^{+}_{\ell'} \quad I^{(3)}_{\ell\ell'} = \int_{-1}^{1/2}d(\cos{\theta})D^{-}_{\ell}D^{-}_{\ell'}; \\  
    &I^{(2)}_{\ell\ell'} = \int_{-1}^{1/2}d(\cos{\theta})D^{+}_{\ell}D^{-}_{\ell'} \quad I^{(4)}_{\ell\ell'} = \int_{-1}^{1/2}d(\cos{\theta})D^{-}_{\ell}D^{+}_{\ell'}.
    \end{split}\label{eq:spin2matrix}
\end{equation}
We write now $D^{\pm}_{\ell}$ in terms of the reduced Wigner rotation matrices, $d^{\ell}_{2,\pm2}$:
\begin{equation}
    D^{\pm}_{\ell}(\cos{\theta}) \equiv [d^{\ell}_{2,2}(\theta)\pm d^{\ell}_{2,-2}(\theta)]
\end{equation}
and define 
\begin{equation}
    I^{\pm\pm}_{\ell\ell'} \equiv \int_{-1}^{1/2}dx \, d^{\ell}_{2,\pm2}(x)d^{\ell'}_{2,\pm2}(x),
\end{equation}
with $x = \cos(\theta)$. The final expression for the $I_{\ell \ell'}^{(X)}$ matrices is then given by
\begin{equation}
     \begin{split}
    &I^{(1)}_{\ell\ell'} =  I^{++}_{\ell\ell'}+I^{+-}_{\ell\ell'}+I^{-+}_{\ell\ell'}+I^{--}_{\ell\ell'}; \\
    &I^{(2)}_{\ell\ell'} =  I^{++}_{\ell\ell'}-I^{+-}_{\ell\ell'}+I^{-+}_{\ell\ell'}-I^{--}_{\ell\ell'};\\  
    &I^{(3)}_{\ell\ell'} =  I^{++}_{\ell\ell'}-I^{+-}_{\ell\ell'}-I^{-+}_{\ell\ell'}+I^{--}_{\ell\ell'}; \\
    &I^{(4)}_{\ell\ell'} =  I^{++}_{\ell\ell'}+I^{+-}_{\ell\ell'}-I^{-+}_{\ell\ell'}-I^{--}_{\ell\ell'}.
    \end{split}
\end{equation}
The matrices $I^{\pm\pm}_{\ell\ell'}$ can be calculated from the relation between Wigner matrices and Clebsch-Gordan coefficients \cite{Varshalovich:1988ye}, as shown in Appendix A of \cite{Yoho:2015bla}.

\end{document}